\documentclass[ba,reqno,preprint]{imsart}
\RequirePackage{amsthm,amsmath,amssymb}
\usepackage{natbib}
\usepackage[colorlinks,citecolor=blue,urlcolor=blue,filecolor=blue,backref=page]{hyperref}
\usepackage{graphicx}
\usepackage{multicol}
\usepackage{newtxmath}
\usepackage{bm}
\usepackage{xfrac}

\arxiv{arXiv:0000.0000}

\startlocaldefs
\numberwithin{equation}{section}
\theoremstyle{plain}
\numberwithin{equation}{section}

\newcommand{\indep}{\perp \!\!\! \perp}

\newcommand{\bbeta}{\mathbf{\beta}}

\newcommand{\bA}{\mathbf{A}}
\newcommand{\bX}{\mathbf{X}}

\newcommand{\bS}{\mathbf{S}}
\newcommand{\SSigma}{\mathbf{\Sigma}}
\newcommand{\ggamma}{\mathbf{\gamma}}
\newcommand{\Oomega}{\mathbf{\Omega}}
\newcommand{\PPsi}{\mathbf{\Psi}}
\newcommand{\mmu}{\mathbf{\mu}}

\newcommand{\mcG}{\mathcal{G}}
\newcommand{\mcM}{\mathcal{M}}

\def\argmin{\mathop{\rm argmin}}
\def\argmax{\mathop{\rm argmax}}
\endlocaldefs

\begin{document}
	
	\begin{frontmatter}
		
		\title{Fast Bayesian High-Dimensional Gaussian Graphical Model Estimation}
		\runtitle{Graphical-SUBHO}

		\begin{aug}
			\author{\fnms{Sagnik} \snm{Bhadury}\thanksref{addr1,t1}\ead[label=e1]{bhadury@umich.edu}},
			\author{\fnms{Riten} \snm{Mitra}\thanksref{addr2,t2}%
				\ead[label=e3]{ritendranath.mitra@louisville.edu}}
			\and
			\author{\fnms{Jeremy} T.\ \snm{Gaskins}\thanksref{addr2,t2}\ead[label=e2]{jeremy.gaskins@louisville.edu}}

			\runauthor{Bhadury et al.}
			
			\address[addr1]{Department of Computational Medicine and Bioinformatics, University of Michigan, $100$ Washtenaw Ave, Palmer Commons Building, Room $2035$, Ann Arbor, MI $48103$, USA
				\printead*{e1}
			}
			
			\address[addr2]{Department of Bioinformatics and Biostatistics, University of Louisville, $485$ E Gray St, Louisville, KY $40202$, USA
				\printead{e3}
				\printead{e2}}
		\end{aug}

			\begin{abstract}
			Graphical models describe associations between variables through the notion of conditional independence. Gaussian graphical models are a widely used class of such models where the relationships are formalized by non-null entries of the precision matrix. However, in high dimensional cases, standard covariance estimates are typically unstable. Moreover, it is natural to expect only a few significant associations to be present in many realistic applications. This necessitates the injection of sparsity techniques into the estimation. Classical frequentist methods use penalization  for this purpose; in contrast, fully Bayesian methods are computationally slow, typically requiring iterative sampling over a quadratic number of parameters in a space constrained by positive definiteness. We propose a Bayesian graph estimation method based on an ensemble of Bayesian neighborhood regressions. An attractive feature of our methods is the ability for easy parallelization across separate graphical neighborhoods, invoking computational efficiency greater than most existing methods. Our strategy induces sparsity with a Horseshoe shrinkage prior and includes a novel variable selection step based on the marginal likelihood from the predictors ranks. Our method appropriately combines the estimated regression coefficients to produce a graph estimate and a matrix of partial correlation estimates for inference. Performance of various methods are assessed using measures like FDR and TPR. Competitive performance across a variety of cases is demonstrated through extensive simulations. Lastly, we apply these methods to investigate the dependence structure across genetic expressions for women with triple negative breast cancer.
		\end{abstract}
		
		\vspace{0.2 in}
		
		
	\end{frontmatter}
	

	\section{Introduction}

	A  graphical model $\mathcal{G}=(V,E)$ can be used to represent the conditional independence structure for multivariate data. Here, $V$ denotes the set of random variables, while the set $E$ is a collection of paired tuples from $V$. $V$ and $E$ can together be represented pictorially as nodes and edges in a connected graph. The graphical representation directly corresponds to the conditional independence structure of the multivariate distribution by the following rule: a pair $(a,b)$ is contained in the edge set $E$ if and only if $ {X}_{a}$ is conditionally dependent on $ {X}_{b}$, given all remaining variables $ {\bX}_{V \setminus \{a,b\}}=\{ {X}_{k};k\in V\setminus \{a,b\}\} $. There can be several  forms of graphical models, allowing for many choices for the marginal density of the node variables.

	If data emerges from a multivariate normal distribution, the graphical model is referred to as the Gaussian graphical model (GGM). The random variables in a GGM follow  a $p$-dimensional multivariate normal distribution with mean $\mmu$ and non-singular covariance matrix $\SSigma$, i.e.,  $\bX=( {X}_{1},\dots,{X}_{p})\sim\mathcal{N}_{p}(\mmu,\SSigma)$. We denote the elements of  the  inverse covariance matrix $\SSigma^{-1} = \Oomega$  as $\omega_{ab}$ (partial covariances) and $\omega_{aa}$ (precisions). Specifically, $\omega_{ab}=0$ implies $(a,b)\notin E$, further showing $ X_a \indep X_b \mid {X}_{V \setminus \{a,b\}}$. That is,  conditional independence between a pair of random variables $a$ and $b$  is equivalent to a zero in the  entries of the inverse covariance. The partial correlation matrix $\PPsi= {(\psi_{ab})}$  may be of additional interest in situations where magnitudes and directions of associations are relevant. The partial correlation between $X_a$ and $X_b$ given rest of the variables ${X}_{V \setminus \{a,b\}}$ is denoted by $\psi_{ab}= - \frac{\omega_{ab}}{\sqrt{\omega_{aa}\omega_{bb}}}$. As in the inverse covariance, zeros in the partial correlation matrix code for conditional independence, and the problem of graph estimation is equivalent to identifying  zeros in $\Oomega$ or in $\PPsi$.

	Hence, estimating a GGM is intrinsically linked with the problem of covariance estimation. 	When $n\gg p$, the classical MLE can provide an accurate estimate for $\SSigma$ \citep{dempster1972covariance}, and the edge estimation (sometimes called covariance selection) is done by considering (approximate) zeros of $\omega_{ab}$. But in high-dimensional cases, the MLE does not exist or may be unstable. In the frequentist domain, this challenge is usually tackled through penalization techniques, whereas Bayesian approaches typically involve sparsity-inducing priors on $\Oomega$.

	A number of frequentist methods perform sparse inverse covariance estimation by 
	optimizing penalized objective functions
	\citep[e.g.,][]{yuan2007model,banerjee2008model}. Two such notable methods for high dimensional inverse covariance estimation  are the Graphical Lasso \cite[GLASSO;][]{friedman2008sparse} and the network exploration  method \citep{fan2009network}. They use the Lasso and the SCAD penalty respectively. The GLASSO is fast, but as noted in \cite{mazumder2012graphical}, it may also fail to converge occasionally. Additionally, \cite{hsieh2013big} proposed an improved algorithm for solving the $\ell_1$-penalized problem in a high-dimensional setting where the number of nodes is on the order of millions.
	
	The  Bayesian literature on covariance and inverse covariance estimation is  rich and constantly growing. \cite{banerjee2014posterior} approached sparse estimation by imposing banding structures on the inverse covariance matrix along with G-Wishart priors. The Graphical Horseshoe \cite[GHS;][]{li2019graphical}, on the other hand, induces sparsity for the precision matrix using element-wise shrinkage priors. Specifically, they use  Horseshoe priors on the off-diagonal elements of $\Oomega$ and an uninformative prior on its diagonal. This use of shrinkage priors directly on the precision entries is similar to the strategy used in the Bayesian Graphical Lasso \citep{wang2012bayesian}, except the latter uses double exponential priors for the off-diagonal elements. There is also an extensive collection of Bayesian strategies built on the Hyper inverse Wishart (HIW) prior structure \citep{dawid1993hyper, roverato2002hyper,carvalho2007simulation,lenkoski2011computational}.  These methods directly specify a prior on the graph space $\mcG$ and  a conditional prior for $\SSigma$ such that the corresponding $\SSigma^{-1}=\Oomega$ will respect the zero structure in $\mcG$.  Despite the many theoretical strengths of the HIW strategy, these methods are often computationally slow because the Markov chain Monte Carlo (MCMC) scheme needs to traverse the discrete space of all possible graphs $\mcG$.

	As an alternative to directly estimating the covariance matrix, \cite{meinshausen2006high} propose to estimate the graph by considering a neighborhood selection (Nbd-Sel) procedure.  This approach exploits the conditional independence structure of GGMs and consecutively performs node-wise penalized regressions. Here, each node is regressed against the remaining nodes as predictors to find the nodes. The final edge set is assembled by combing the results from  all node-wise regressions. In an analogous setting, \cite{peng2009partial} assumed an overall sparsity of the partial correlation matrix and employed sparse regressions for model fitting. TIGER by \cite{liu2017tiger}, an asymptotically tuning-free and non-asymptotically tuning-insensitive method for high-dimensional GGM, is an extension of the Nbd-Sel approach, where the LASSO estimator is replaced with a square-root LASSO estimator \citep{belloni2011square}. Another approach by \cite{cai2011constrained} employs the Dantzig selector for solving each node-wise sparse regressions through the CLIME algorthm. A quasi-Bayesian implementation of the Nbd-Sel method was developed by \cite{atchade2019quasi}, producing an efficiently calculable product-form quasi-posterior distribution. Our proposed methodology  is also closely aligned with the  Nbd-Sel  framework and  is substantially motivated by computational considerations. As Nbd-Sel treats the graph estimation problem as a collection of regressions, computations can be efficiently performed through  the use of parallelization.

	An outline of the article is as follows. First, we establish the nomenclature associated with the Nbd-Sel and GGM in Section~\ref{sc: Neighborhood Selection}. We further describe the relationship between multiple regressions and the elements of $\Oomega$. In Section~\ref{sc: Bayesian framework for Variable Selection}, we propose a new variable selection strategy for performing edge selection in a Bayesian model with a Horseshoe prior. We introduce our Step-up Bayesian Horseshoe method (SUBHO), which combines posterior inference from continuous shrinkage priors with a forward edge selection algorithm for neighborhood estimation. In Section~\ref{ssc:Comparisons}, we discuss a host of other related methods that use alternative  priors for sparsity control. Some of these methods are novel variants of our general approach that combine the structure of neighborhood selection along with other sparsity priors traditionally used in Bayesian regression (Spike-and-slab and Hyper-$g$). In Section~\ref{sc: numericals}, we perform multiple numerical experiments to evaluate and compare the recovery of the underlying graph structure and the partial correlations. In Section~\ref{sc: TNBC data analysis}, we demonstrate  our approach by estimating the dependence structure in  genetic expression data for triple negative breast cancer patients from the BRCA-METABRIC study. Finally, we end by discussing limitations and future directions.

	\section{Neighborhood Selection and GGM}\label{sc: Neighborhood Selection}
	Following the usual graph theory terminology, we denote the neighborhood $ne_a$ of a node $a \in V$  to be the subset of $V \setminus \{a\}$ such that $ne_a \subset V$ consists of all nodes $b \in V \setminus \{a\} $ such that $(a, b) \in E$. Nbd-Sel is a divide-and-conquer approach to covariance selection that relies on estimating these individual neighborhoods for every node. Such an approach has been utilized in multiple frequentist procedures for graphical estimation \citep[e.g.,][among others]{meinshausen2006high,cai2011constrained,liu2017tiger}. Estimating a neighborhood for a particular node $a$ translates to finding its relationship to the remaining nodes $V \setminus \{a\}$. This can be  framed  characterizing the conditional distribution of $X_a$ given  $\bX_B=\{\bX_b:b\in V\setminus\{a\}\}$. Estimating this conditional distribution is typically treated  as regressing the response $X_a$ against the predictor variables $\bX_B$. The significant/selected variables obtained from this regression  determine the neighborhood of the node $a$. Once the sets $ne_a$ are estimated for each  $a$, they are combined to estimate the edge set $E$ and its corresponding graph $\mcG$.

	To investigate the relationships between the parameters associated with the node-specific regression models and the partial correlations, we recall that the joint density (implied by the GGM) can be expressed as $\bX=( {X}_{1},{X}_{2},...,{X}_{p})\sim\mathcal{N}_{p}(\mmu,\SSigma)$, with mean $\mmu$ and non-singular covariance matrix $\SSigma=\Oomega^{-1}$. For  a fixed node $a\in V$, we consider the conditional distribution of ${X}_a$ given the remaining $(p-1)$ variables in $B = \{V\setminus\{a\}\}$, i.e., $ \bX_{B} = ( {X}_1,\ldots, {X}_{a-1}, {X}_{a+1},\ldots, {X}_{p})$. Relying on standard multivariate normal results, this distribution is given by 
	\begin{eqnarray}
		{X}_a\mid \bX_{B} &\sim& \mathcal{N}(\mu_a+\SSigma_{aB}\SSigma_{BB}^{-1}(\bX_{B}-\mmu_{B}) , \Sigma_{aa}-\SSigma_{aB}\left(\SSigma_{BB}\right)^{-1}\SSigma_{Ba}),
		\label{eq:conditional rep of MVTnorm 1}
	\end{eqnarray}
	where the joint covariance of $\bX=({X}_a, \bX_{B})$ is a block matrix with $\Sigma_{aa}$ and $\SSigma_{BB}$ denoting the covariance of ${X}_a$ and $\bX_{B}$, respectively, and $\SSigma_{aB}$ is a $(p-1)$ vector of covariances between the two. The form of the conditional distribution in Equation~\eqref{eq:conditional rep of MVTnorm 1} resembles the following regression model 
	\begin{eqnarray}
		X_{a} &=& \beta_{a}^{(0)} + \bX_{B} \bbeta_{a} + \epsilon_a.
		\label{eq: intercept part out of beta 1}
	\end{eqnarray}
	Mapping  to the terms in Equation~\eqref{eq:conditional rep of MVTnorm 1}, 
	note that the intercept is $\beta_{a}^{(0)}= \mu_a - \SSigma_{aB}\SSigma_{BB}^{-1}\mmu_{B}$\\ 
	$= \mu_a + \Oomega_{aB} \Omega_{aa}^{-1}\mmu_{B} $ and the regression coefficients are $\bbeta_{a} = \SSigma_{aB}\SSigma_{BB}^{-1} = -\Oomega_{aB} \Omega_{aa}^{-1}$. We define the matrix $\bbeta$ such that the $(a,b)$ element  is $\beta_{a}^{(b)}$, the coefficient of the predictor $X_b$ when $X_a$ is the response variable. This is, the $b^{th}$ element of $\bbeta_{a}$. From $\bbeta_{a} =  -\Oomega_{aB} \Omega_{aa}^{-1}$, we have that $\beta_{a}^{(b)} = -{\omega_{ab}}/{\omega_{aa}}=\psi_{ab} \sqrt{{\omega_{bb}}/{\omega_{aa}}}$, which implies that $\Oomega,\PPsi$, and  $\bbeta$ all have the same sparse structure. That is, if $(a,b)\notin E$, then $\beta_a^{(b)}=\beta_b^{(a)}=\omega_{ab}=\psi_{ab}=0$.

	We have shown that one can obtain   $\bbeta_a$, the parameters associated with conditional distributional of each node, directly from $\Oomega$. Exploring the opposite direction, we can also obtain 
	the partial correlation matrix $\PPsi$ from the node-wise regressions by inverting the above relationship. 
	From $\beta_{a}^{(b)}\beta_{b}^{(a)} = {\omega_{ab}^2}/{\omega_{aa}\omega_{bb}}=\psi^2_{ab}$ and $\text{sign}(\psi_{ab})=\text{sign}(\beta_{a}^{(b)})=\text{sign}(\beta_{b}^{(a)})$, it follows that 
			\begin{eqnarray}
			\label{eq: par cor from beta 1}
			\psi_{ab} = \text{sign}(\beta_{a}^{(b)}) \sqrt{ \beta_{a}^{(b)}  \beta_{b}^{(a)}}. 
		\end{eqnarray}
	Here, $\text{sign}(a) = {a}/{|a|}$ for $a \neq 0$ and is defined to be  zero if $a=0$. Importantly, this suggests that the zero structure of the partial correlation matrix can be learned from the zeros in $\bbeta$. Motivated by this, we  design  our Bayesian graphical inference  to be a combination of Bayesian regressions for each node $a$. For computational stability we center and standardize all columns of $\bX$ before executing the regressions. For notational convenience, we include a column of ones in $\bX_{B}$ such that $\bbeta_a$ automatically includes the intercept, although we need not keep track of $\beta_a^{(0)}$ in the matrix $\bbeta$. We can execute each regression in parallel across multiple CPU/GNU cores. Each regression provides  an estimate of the neighborhood for a given node $(\hat{ne}_a)$, which can then be combined to obtain $\hat{\mcG}$. Additionally, $\hat{\bbeta}_a$s can be used to estimate $\hat{\PPsi}$ in accordance with Equation~\eqref{eq: par cor from beta 1}. Unlike frequentist Nbd-sel approaches, our Bayesian framework for regression and variable selection needs to be defined by appropriate priors on the regression coefficients. Since our aim is sparse graph estimation, we will focus on priors that encourage sparsity in node-wise regressions. Such priors will be  the focus in the subsequent section.

		\section{Graph Estimation with Step-Up Bayesian model selection for HOrseshoe (SUBHO)}\label{sc: Bayesian framework for Variable Selection}

		Having translated the problem of graphical estimation into an ensemble of regressions models of the form ${X}_a = \bX_{B}  \bbeta_a + \epsilon_a$, we now seek an approach that would encourage sparse estimates for $\bbeta_a$.	As noted previously, many frequentist solutions to this problem have been  considered \citep{meinshausen2006high,yuan2007model,banerjee2008model,liu2017tiger,cai2011constrained}.	We instead consider a Bayesian variable selection approach for this step.

		In a Bayesian context, one of the primary approaches for variable selection is based on using mixture priors for the regression coefficients. In its typical form, the spike-and-slab approach considers an independent mixture for each regression coefficient including a zero or narrow spike prior for when the predictor is irrelevant and a diffused slab prior for large effects
		\citep{mitchell1988bayesian, george1993variable}. A related approach is to consider a mixture of $g$-priors to introduce correlation among the selected predictors
		\citep{fernandez2001benchmark, liang2008mixtures, bove2011hyper, li2018mixtures}. In both cases, the selected variables are typically chosen by the median model (i.e., the model containing the predictors with posterior inclusion probability exceeding 50\%), as advocated by \cite{barbieri2004optimal}.  Coefficient estimation follows from using Bayesian Model Averaging (BMA) or from the selected median model.  As these methods search over a discrete model space, Markov chain Monte Carlo and other computational algorithms typically scale poorly to high dimensions.

		Shrinkage priors have recently gained popularity as a more computationally efficient alternative. In contrast to selection priors, shrinkage priors are fully continuous distributions and use a prior density with high peak around zero to pull or ``shrink'' small signals aggressively towards zero, while also including a thick tail to accommodate  large non-zero effects. Most shrinkage priors fall within the family of global-local (GL) shrinkage priors which have the form $\beta_i\sim \mathcal{N}(0,\tau^2 \lambda_i^2)$, $\lambda_i \sim f(\cdot)$, $\tau \sim g(\cdot)$ \citep{polson2010shrink}. Here, $\tau$ controls global shrinkage towards the origin, while the local shrinkage parameters $\lambda_1,\ldots,\lambda_p$ allow differences in the degree of shrinkage between individual predictors. Typically, $f(\cdot)$ and $g(\cdot)$ are chosen to have heavy tails and substantial mass near zero, respectively. The Normal-Gamma  \citep{brown2010inference} and Horseshoe \citep{carvalho2010Horseshoe}
		priors are the most commonly used shrinkage priors.	Compared to selection priors, a main difficulty to using shrinkage is that the posterior probability of attaining an exact value zero for the coefficient is always zero. Thus, the sparsity structure  cannot be directly obtained from the MCMC output.

		\subsection{Horseshoe Shrinkage: Estimation of $\beta_a$}\label{ssc: GL Horseshoe framework}
		
		The  Horseshoe prior \citep[HS;][]{carvalho2010Horseshoe} for shrinkage is $\beta_{a}^{(b)}\sim \mathcal{N}(0,{\lambda_{a}^{(b)2}}\tau_a^2\sigma_a^2)$ where $\lambda_{a}^{(b)} \sim C^+(0,1)$, $\tau_{a} \sim C^+(0,1)$ for $b=0,1,\ldots,p$; $b\ne a$. Here, $C^+(0,1)$ is the standard half-Cauchy distribution, taken as the prior for both the local parameters $\lambda_{a}^{(b)}$ and for the  global parameter $\tau_a$. The $\sigma^2_a$ term is the  residual variance for the regression model and also acts as the scale for the regression coefficient prior. The global hyperparameter provides shrinkage to all $\beta_{a}^{(b)}$s, whereas the local hyperparameters allow stronger signals to avoid shrinkage, thus encouraging the identification of a few non-zero relationships. Posterior inference using Gibbs sampling can be performed using the popular data augmentation technique from \cite{makalic2015simple} that re-characterizes the square of the $C^+(0,1)$ as a mixture of inverse gammas. For the residual variance $\sigma_a^2$, we use a dispersed $\mathcal{IG} (1/10,1/10)$ prior. Sampling from this posterior is straight-forward, and the conditional distribution for the regression coefficients is	$\bbeta_a\mid\cdots \sim \mathcal{N}_p(\bA_a^{-1}\bX_{B}^{T} X_{a},\sigma_a^2\bA_a^{-1})$ where $\bA_a  =  (\bX_{B}^{T}\bX_{B}+(\mathbf{\Lambda}_{a}^{*})^{-1}) $ and $\mathbf{\Lambda}_{a}^{*}$ is a diagonal matrix with elements  $\tau_a^2\lambda_{a}^{(b)2}$, $(b=0,1,\ldots,p; b\ne a)$. Note that inverting the matrix $\bA_{a}$ can pose a computational bottleneck in high dimensional settings. In such cases, we follow the algorithm from \cite{bhattacharya2016fast} that exploits the scale-mixture representation of our samplers.

		\subsection{Step-Up Bayesian model selection for HOrseshoe (SUBHO)}\label{sssc: SWHS selection}

		Recall that our goal is to estimate the full graph structure by individually estimating the neighborhoods for node $a$ ($a=1,\ldots,p$). From an individual HS regression, we obtain the point estimate $\hat{\bbeta}_a$ by averaging the posterior samples, but $\hat{\bbeta}_a$ will never contain exact zeros. Thus, one cannot directly use this to determine which variables are significant predictors (i.e., neighbors) of node $a$. There are multiple ways of selecting predictors from a specific HS model. As suggested by a number of authors \citep{carvalho2010Horseshoe,van2017adaptive,van2017uncertainty}, one can either check whether zero is contained in a user specified credible set ($50\%$ or $95\%$) to decide if a feature is significantly associated. Alternatively, one could apply some thresholding rules to the posterior estimates. In preliminary simulations we  found both these approaches to be ineffective, leading us to propose a strategy which we call Step-Up Bayesian model selection for HOrseshoe (SUBHO).

		Notationally, let $\ggamma_a=(\gamma_a^{(1)},\ldots,\gamma_a^{(a-1)},\gamma_a^{(a+1)},\ldots,\gamma_a^{(p)})$ denote the set of variable selection indicators. Here $\gamma_a^{(b)} = 1$ indicates that $X_b$ is included in the set of predictor variables  for outcome node $X_a$, and $\gamma_a^{(b)} = 0$ indicates exclusion of $X_b$. The number of active predictors under a particular $\ggamma_a$ configuration is given by $p_{\ggamma_a} = \sum_b \gamma_a^{(b)}$. Each $\ggamma_a$ implies a particular model among the $2^{p-1}$ potential models for the neighborhood of node $a$. Our strategy is to aggressively reduce this model space using the point estimate from the HS model to avoid the overwhelming burden of calculating all possible potential models. We index the models by $\ggamma_a$ and denote the model space for all possible potential candidate models as $\mcM_{\ggamma_a}$. For a given $\ggamma_a$ vector, the regression model becomes 
		\begin{eqnarray}
			\mcM_{\ggamma_a}: \space X_{a} 
			&=& \beta_{a}^{(0)} + \bX_{B}^{(\ggamma_a)}\bbeta_{a}^{(\ggamma_a)} + \epsilon, \label{eq: model space}
		\end{eqnarray}
		where $\bX_{B}^{(\ggamma_a)}$ represents the $(n \times p_{\ggamma_a})$ design matrix under model $\mcM_{\ggamma_a}$, and $\bbeta_{a}^{(\ggamma_a)}$ is the $p_{\ggamma_a}$-dimensional vector of nonzero regression coefficients. We note that the intercept  $\beta_a^{(0)}$ is always included in each of our models ($\gamma_a^{(0)}=1$ with probability $1$). From the MCMC samples from the HS model with all predictors, we compute the posterior mean $\hat{\bbeta}_a = ( \hat{\beta}_{a}^{(b)})_{b\ne a}$. As the data has been standardized before running MCMC, these estimates represent a sparse estimate of the partial regression coefficients, and we may take the magnitude as a marker of the importance of $b^{th}$ predictor. To that end, we define the mapping $R(\cdot):\{1,\ldots,p-1\}\rightarrow \{V\setminus\{a\}\}$ as a ranking function of the absolute value of the estimated coefficients (excluding the intercept), i.e., $R(1)=\argmax_{b}{\mid\hat{\beta}_{a}^{(b)}\mid}$ provides the predictor with the largest estimated coefficient, $R(2)$ designates the predictor with the second largest estimated coefficient, etc. Rather than investigating all $2^{p-1}$ combinations of $\ggamma_a$, we consider models that have the following form: $\gamma_a^{(b)}=1$ if $b\in\{R(1),\ldots,R(k)\}$ and $\gamma_a^{(b)}=0$ if $b\in\{R(k+1),\ldots,R(p-1)\}$. That is, we only allow the $k$-highest ranked predictors to be in the model for $k=0,\ldots,K$. We define the $k=0$ model to exclude all predictors, ($\gamma_a^{(b)}=0$ for all $b\ne 0$), and $K$ $(\le p-1)$ represents the number of predictors allowed in the densest model considered. We denote this restricted candidate model space by $\widetilde{\mcM}_{\ggamma_a}$. For each $\ggamma_a$ in our restricted model space $\widetilde{\mcM}_{\ggamma_a}$, we compute the marginal likelihood for  this model by 
		\begin{equation} p(X_{a}|\ggamma_{a}) = \int  \int
			p(X_{a}|\bbeta_{a}^{(\ggamma_{a})},\sigma^2,\bX_{B}^{(\ggamma_{a})})p(\bbeta_{a}^{(\ggamma_{a})},\sigma^2) \,d\bbeta_{a}^{(\ggamma_{a})}\,d\sigma^2, \nonumber
		\end{equation}
		where we restrict the design matrix to only those $\ggamma_{a}$ in the restricted model space $\widetilde{\mcM}_{\ggamma_a}$. However, we cannot compute this integral in closed form due to the non-conjugacy of the HS prior structure. Instead, we propose a simpler working model for computing these marginal likelihoods based on the simple conjugate prior, 
		\begin{equation}
			p_W(\bbeta_{a}^{(\ggamma_{a})}|\sigma_{a}^2 ) = \mathcal{N}(0,  \sigma_{a}^2 I_p) \text{  and  } p_W(\sigma_{a}^2) = \mathcal{IG}(c,d), 
			\label{eq: working mdoel prior}
		\end{equation}
		where typically $c=d=1$. Note that the working prior distribution in Equation~\eqref{eq: working mdoel prior} facilitates the closed-form calculations of the marginal likelihood for each $\ggamma_a \in \widetilde{\mcM}_{\ggamma_a}$. It is clear that this working prior does not encourage any variable shrinkage among the $k$ retained coefficients, but this is acceptable as we are only using it for our $k$-rank models, which is already sparse in the sense that the other $p-1-k$ coefficients are zero. Since we typically limit ourselves to small values of $k$ ($K$ is usually much less than $p-1$), the lack of variable shrinkage in the working model is not a concern. Hence, our primary prior still remains the HS prior which performs the key role of ranking the predictors and constructing the model space. Under this simpler working model, the marginal likelihood is 
		\begin{eqnarray}\label{eq: working model joint final}
			p({X}_a \mid {\ggamma_a}) &=& 
			\int \int 
			p({X}_a\mid\bbeta_{a}^{(\ggamma_a)},\sigma^2,\bX_{B}^{(\ggamma_a)})p_W(\bbeta_{a}^{(\ggamma_a)},\sigma^2) \,d\bbeta_{a}^{(\ggamma_a)} \,d\sigma^2 \nonumber \\
			&=& \frac{d^c \,\Gamma(c+n/2)}{(2\pi)^{n/2} \,\Gamma(c)} \left(\mid V^*\mid\right)^{\sfrac{1}{2}} (d^*)^{-c^*} 
		\end{eqnarray}
		where $\mathbf{V}^* =
		(\mathbf{I}_{p_{\ggamma_a}}+(\bX_{B}^{(\ggamma_a)})^{T} {\bX_{B}^{(\ggamma_a)} })^{-1}$, $ \mmu^* = \mathbf{V}^*(\bX_{B}^{(\ggamma_a)})^{T} X_{a}$, $c^*=  c + n/2$, and $d^* =  d+\frac{1}{2}\{  {X}_{a}^{T}{X}_{a}-(\mmu^*)^{T}(\mathbf{V}^*)^{-1}\mmu^*\} $. 
		
		To complete the SUBHO model selection step, we assign a Beta-Bernoulli prior on the $\ggamma_a$ from the hierarchy $\xi \sim Beta(\alpha^*,\beta^*)$ and $\ggamma_a \mid \xi \sim_{iid} Bern(\xi)$. Marginally the prior is
		\begin{eqnarray}\label{eq:prior on model}
			p(\ggamma_a) = \int p(\ggamma_a \mid\xi)p(\xi)\,d\xi = \frac{\Gamma(\alpha^* + \beta^*)}{\Gamma(\alpha^*) \Gamma(\beta^*)}  \frac{\Gamma(\alpha^* + p_{\ggamma_a})  \Gamma((p-1)+\beta^*-p_{\ggamma_a})}{\Gamma(\alpha^*+\beta^*+(p-1))} .
		\end{eqnarray}
		Note that as $\gamma_a^{(0)}$ is fixed as one (intercept always included), only the remaining $p-1$ components of $\ggamma_a$ are modeled through this prior. Finally we get the posterior probability from 
		$p(\ggamma_a \mid X_a) \propto p(X_a\mid\ggamma_a)\times p(\ggamma_a) $. For the fixed $a^{th}$ node wise regression, we select that particular model which has the maximum posterior probability for $\ggamma_a$ over the rank-restricted space $\widetilde{\mcM}_{\ggamma_a}$, $\hat{\ggamma}_{a} = \argmax_{\ggamma_a \in \widetilde{\mathcal{M}}_{\ggamma_a}}p(\ggamma_a \mid X_a) $.

		\subsection{Combining the Results}\label{ssc:postproc 1}

		Implementing our strategy involves fitting $p$ separate regression models with the HS prior and performing the Step-Up Bayesian model selection to get the neighborhood for each node. Following this procedure, we get $\hat{\bbeta}_a$ and $\hat{\ggamma}_a$ for all $a=1,\ldots,p$. Note that these $\hat{\bbeta}_a$ estimates are  a sparse (but non-zero)  set of coefficients from individual Horseshoe regressions, whereas the variable inclusion indicators from SUBHO are $\hat{\ggamma}_a$. These results must be combined in a suitable way to obtain the estimates of the parameters of interest---the graph $\mcG$ and the partial correlation matrix $\PPsi$. To this end, we collect the results across all node-wise regressions to obtain $\hat{\bbeta}_{all}$ and $\hat{\ggamma}_{all}$, where the $(a,b)^{th}$ elements 
		are  $\hat{\beta}_a^{(b)}$ and $\hat{\gamma}_a^{(b)}$, respectively.

		\subsubsection{The Graph}\label{sssec: Graph. Mtx.}
		
		To develop an estimator for the graph $\mcG$ using $\hat{\ggamma}_{all}$, note that $\hat{\ggamma}_{all}$ will not generally be symmetric as each neighborhood is estimated in parallel without sharing information. A symmetrized version is necessary to produce a reasonable estimate of the graph. Following \cite{meinshausen2006high}, we consider the AND $(\land)$ and OR $(\lor)$ rules for symmetrization: 
		\begin{eqnarray}
			\hat{E}^{\land}= \{(a,b): \hat{\gamma}_a^{(b)}=\hat{\gamma}_b^{(a)}=1\} \text{  and  } \hat{E}^{\lor}= \{(a,b): \hat{\gamma}_a^{(b)}=1 \text{ or } \hat{\gamma}_b^{(a)}=1\} .
			\label{eq: symmetrization}
		\end{eqnarray}
		In words, the AND edge set for $\hat{\mcG}=(V,\hat{E}^{\land})$ contains only those $(a,b)$ where $a$ is included in the estimated neighborhood of $b$ and vice versa. The OR edge set $\hat{E}^{\lor}$, on the other hand, includes those  $(a,b)$ which satisfy the inclusion criterion in at least one direction. This is a more liberal edge-selection rule and will necessarily produce denser graphs. In another related approach, \cite{peng2009partial} proposed a joint sparse regression based on neighborhood regressions. They also minimize an $\ell_1$ penalized loss function, but their joint sparse regression model simultaneously performs  Nbd-Sel for all nodes together---instead of examining each separately in parallel---and thus, is computationally more burdensome. The benefit to this choice is that they guarantee symmetrization by forcing $\hat{\gamma}_a^{(b)} = \hat{\gamma}_b^{(a)}$.

		\subsubsection{Partial Correlation Matrix}\label{sssec:partial Cor. Mtx.}
		
		We propose an estimate of the partial correlation matrix based on the relation between $\PPsi$ and $\bbeta$ from Equation~\eqref{eq: par cor from beta 1}. Specifically, we obtain $\hat{\PPsi}=(\hat\psi_{ab})$ from $\hat{\bbeta}_{all}$ by
		\begin{eqnarray}
			\hat\psi_{ab}= 
			\begin{cases}
				\text{sign}\left(\hat{\beta}_{a}^{(b)}\right)  \left[\sqrt{\hat{\beta}_{a}^{(b)} \hat{\beta}_{b}^{(a)}}\right]_{\leq 1}, \ \    &\text{sign}\left(\hat{\beta}_{a}^{(b)}\right) = \text{sign}\left(\hat{\beta}_{b}^{(a)}\right), \\
				0,   &\text{sign}\left(\hat{\beta}_{a}^{(b)}\right) \ne \text{sign}\left(\hat{\beta}_{b}^{(a)}\right), \\
			\end{cases}
			\label{eq: par cor adj}
		\end{eqnarray}
		where $[x]_{\leq 1} = xI(x\leq 1) +I(x> 1)$. Recall that Equation~\eqref{eq: par cor from beta 1} guarantees that the true values of ${\beta}_{b}^{(a)}$ and ${\beta}_{a}^{(b)}$  have the same sign, but while fitting the $p$ separate regressions, it may turn out that $\text{sign}(\hat{\beta}_{b}^{(a)})$ differs from $\text{sign}(\hat{\beta}_{a}^{(b)})$. In our simulations, we have seen that this sign discrepancy most frequently occurs when both $\hat{\beta}_{b}^{(a)}$ and $\hat{\beta}_{a}^{(b)}$ are close to zero due to  high shrinkage, and so estimating the partial correlation to be zero in such a case is natural. 	Additionally, since the $\hat\psi_{ab}$s are calculated from the estimates  $\hat{\beta}_{a}^{(b)}$ and $\hat{\beta}_{b}^{(a)}$, there is a possibility that their absolute value might exceed $1$. Though we have not encountered  this scenario in any of our simulations, it could be accounted for by the  truncation function in the above equation.	We should note that we use the shrunk (but non-zero) $\hat{\bbeta}_a$s from the HS output and utilized these estimates to obtain the sparse (but again, non-zero) estimate of $\hat{\PPsi}$. As a final comment, it might be desirable for the estimated partial correlation matrix to be positive definite. In principle, this regression based method will not guarantee the positive definiteness of the resulting estimator. We acknowledge that this is a limitation of our method, but our primary interest lies in recovering the underlying graph structure with minimal plausible error, we are less concerned with the lack of positive definiteness.
		
		\section{Competing Methods}\label{ssc:Comparisons}
		
		In the section we present and discuss a collection of the competitor methodologies and their corresponding estimation procedures for  graph and partial correlation matrix estimation.

		\subsection{GGMProjpred with Horseshoe (Projpred-HS)}\label{sssc: GGMprojpred }
		
		Similar to our approach, \cite{williams2018bayesian} also approach the GGM problem through a Bayesian neighborhood selection framework.  They seek to learn the neighborhood for each node $a$ by utilizing the posterior samples of the regression model using a HS prior on the regression coefficients to induce sparsity. However, the main difference lies  in the model selection step. They consider the projection predictive model selection technique \citep{piironen2020projective},  a decision-theoretic approach based on optimizing predictive performance. An initial reference model $\mcM_*$ is chosen  as a baseline for predictive performance from which  MCMC samples model are obtained. In this context,  $\mcM_*$  is  the regression model with a HS prior on all predictors. $\mcM_{\bot}$ represents a competitor model (i.e., a model with a reduced set of predictors), and a set of  posterior  samples for $\mcM_{\bot}$ are obtained by ``projecting'' the $\mcM_*$ samples  to the $\mcM_{\bot}$ space without running a new MCMC. For a linear model, these KL projection samples can be obtained in closed-form \citep{piironen2017comparison}. Based on the estimated posterior distribution for the competitor model $\mcM_\bot$, the predictive performance between $\mcM_\bot$ and $\mcM_*$ is compared through  an accuracy measure $u=u(\mcM)$. To this end, they recommend the log-predictive density using an approximate leave-one-out cross-validation (LOOCV) with Pareto smoothed importance sampling, to  avoid the repeated fitting of the reference model \citep{vehtari2017practical}. The goal is to select the simplest/sparsest model $\mcM_\bot$ that will lower the predictive performance relative to the reference model $u_*=u(\mcM_*)$ by no more than an acceptable small difference  $\Delta u$  with some level of confidence $\alpha$, i.e., we want to choose the simplest model $\mcM_\bot$ such that $P\{u_* -u(\mcM_\bot) \leq \Delta u\} \leq 1- \alpha$, where this probability is estimated using a Bayesian bootstrap. The set of models $\mcM_\bot$ that are considered  are chosen by a forward selection algorithm.

		\cite{piironen2017comparison} recommend  the default choice for  $\alpha$ to be $0.10$ and for $\Delta u$ to be  $5\%$ of the difference $\hat{u}_* -\hat{u}_0$, where $\hat{u}_0$ denotes the accuracy estimate for the simplest possible (null) model. \cite{williams2018bayesian} made the code for their method available on Github, but due to R library dependencies, we were unable to implement it directly. Instead, we  extracted portions of their code from within their functions to get a working version of their algorithm, but there is a potential that this has  introduced errors into the code. We have found the implementation of the LOOCV step to be  incredibly slow, and this leads the entire method to be quite time-consuming compared to other approaches. From each individual nodal regressions we obtain $\hat{\ggamma}_a$ according to the selected model $\mcM_\bot$, which are combined across all $a$ and the graph $\mcG$ is estimated as described in Section~\ref{ssc:postproc 1}. While fitting the reference model with all predictors and the HS prior, we obtain sparse but non-zero  $\hat{\bbeta}_a$ and estimate $\PPsi$ as in SUBHO.

		\subsection{Graph Estimation with BASAD}\label{ssc: basad}
		
		A key advantage of a Bayesian version of the Nbd-Sel approach is that it is amenable to a wide class of variable selection priors. In addition to the HS shrinkage priors, we can also consider variable selection through spike-and-slab priors. We consider spike-and-slab formulation given by \cite{narisetty2014bayesian} called BAyesian Shrinking And Diffusing priors (BASAD). 
		Consistent with our prior notation, $\ggamma_a = (\gamma_a^{(0)},\gamma_a^{(1)},\ldots,\gamma_a^{(a-1)},\gamma_a^{(a+1)},$ $\ldots,\gamma_a^{(p)})$ denotes variable selection indicators for the  node $a$ regression model. The prior of $\beta_{a}^{(b)}$ under $\gamma_a^{(b)} = 0$ is highly concentrated around zero and referred to as the spike distribution, and the diffuse prior under $\gamma_a^{(b)} = 1$ is called the slab prior. The BASAD model considers the following model: ${X}_a\mid\bX_{B},\bbeta_a,\sigma^2  \sim \mathcal{N}(\bX_{B}\bbeta_a,\sigma^2I_n)$, with the hierarchical prior  $\beta_{a}^{(b)}\mid\sigma^2,\gamma_{a}^{(b)}=0 \sim \mathcal{N}(0,\tau_{0}^2 \sigma^2)$ (concentrated) and  $\beta_{a}^{(b)}\mid\sigma^2,\gamma_{a}^{(b)}=1 \sim \mathcal{N}(0,\tau_{1}^2 \sigma^2)$ (diffuse). Here $ \tau_0\ll \tau_1$ are fixed constants that enforce small variance for the point mass distribution and high dispersion for the diffuse prior component. Following \cite{narisetty2014bayesian}, we take $\tau_0^2 ={\hat{\sigma}^2}/{10n}$ and $\tau_1^2 =\hat{\sigma}^2 \max ( {p^{2.1}}/{100n}, \log n )$ where $\hat{\sigma}^2$ is the sample variance estimator for $X_{a}$. We use  $\sigma^2 \sim \mathcal{IG} (0.1,0.1)$. We always include the intercept in the models, i.e., $\gamma_a^{(0)}=1$. A prior on the model space is encoded through the prior on the $\ggamma_a$; $\gamma_{a}^{(b)}  \sim Bern( q_n )$ where the variable inclusion probability  $q_n $ is chosen to satisfy $P(\sum_{b=1}^{p} \gamma_{a}^{(b)} > K )=0.1 $ for  $K = \max(10,\log(n))$.

		As in \cite{george1993variable}, the proposed sampling scheme for BASAD alternates between sampling $(\bbeta_a \mid\ggamma_a)$ and $(\ggamma_a\mid\bbeta_a)$. By sampling the variable inclusion indicators conditionally on the current coefficient value, this sampler runs much faster than algorithms than sample marginally over $\bbeta_a$, although	this may come with an increased chance of the algorithm getting stuck at a local mode for $\ggamma_a$. We use the posterior probabilities of the latent variables $\ggamma_{a}^{(b)}$ to identify the active covariates. The significant nodes from each node-wise regression are chosen from the median probability model (MPM) to  estimate $\ggamma_a$ \citep{barbieri2004optimal} and the corresponding graph.
		Similarly, $\PPsi$ is estimated using the posterior means of $\bbeta_a$, which are sparse but do not contain exact zeros.

		\subsection{Graph Estimation with Hyper-$g$ prior (BAS)}\label{ssc: hyper-g}
		
		Here, we consider the hyper $g$-prior for variable selection \citep{fernandez2001benchmark, liang2008mixtures, bove2011hyper, li2018mixtures}. For a fixed node $a$, the model/variable inclusion indicators $\ggamma_a$ follow a  Beta-Binomial prior as in \eqref{eq:prior on model}. Conditionally on $(\ggamma_a,g)$ the prior for $\bbeta_a^{(\ggamma_a)}$ (the non-zero elements of $\bbeta_a$) is a mean-zero multivariate normal distribution with covariance matrix $g\sigma_a^2(({\bX_{B}^{(\ggamma_a)}})^{T}\bX_{B}^{(\ggamma_a)})^{-1}$.
		Unlike BASAD, the coefficients with $\gamma_a^{(b)}=0$ take an exact zero value ($\beta_a^{(b)}=0$).
		The scalar hyper-parameter $g>0$ controls both shrinkage toward the prior mean and can be interpreted as an inverse relative prior sample size. Fully Bayesian inference can be conducted with the hyper-$g$ prior \citep{liang2008mixtures}, $p(g) = \frac{a-2}{2}(1+g)^{-{a}/{2}}$, which is a proper distribution with $a>2$. The intercept $\beta_a^{(0)}$ is given an independent locally uniform prior, and the residual variance $\sigma_a^2$ is assigned the Jeffreys' prior.

		Inference with this prior can be done using the Bayesian Adaptive Sampling (BAS) package \citep{BASpackage}. This algorithm selects $\ggamma_a$ without replacement for marginal likelihood/posterior probability calculation using an adaptive strategy where the chance that $\ggamma_a$ is selected to compute $p(X_a\mid\ggamma_a)$ is adaptively updated	based on the previously sampled models and	an efficient MCMC search algorithm which samples models using a tree structure of the model space \citep{clyde2011bayesian}. Rather than evaluating the posterior probability over all $2^{p-1}$ $\ggamma_a$ models, our implementation ensures that each node regression samples no more than $2^{20}$ distinct $\ggamma_a$ for calculation. Further, to avoid search over unreasonably large models, we use a truncated Beta-Binomial prior	that assigns zero probability to all models with $p_{\gamma_a}>K$ where $K=p-1$ and $\alpha=\beta = 1$. The median probability model is considered to obtain $\hat{\ggamma}_a$, and $\hat{\bbeta}_a$ comes from the Bayesian model averaged estimate; $\hat{\mcG}$ and $\hat{\PPsi}$ are obtained by combining results across all nodal regressions as in Section~\ref{ssc:postproc 1}.

		\subsection{Inverse Wishart Prior (IW)}\label{ssc: Inverse Wishart}
		
		In contrast to models using sparse inference on graph structures, the  Inverse Wishart (IW) approach directly infers the  covariance $\SSigma$. IW is among the simplest prior models for covariance estimation and uses the conjugate Inverse Wishart prior $\SSigma \sim IW(m,V)$.  The  hyperparameters $V=I_p$ and $m=p+2$ are interpreted as the prior scale matrix and the degrees of freedom. Assuming the prior mean of $\bX$ is zero (which may be reasonable since data are centered), the posterior distribution of $\SSigma$ is known to be $IW(m+n,V+\bX^T\bX)$. Finally, our estimate for $\hat{\PPsi}$ is obtained by re-scaling the posterior mean for the precision matrix: $\hat{\Oomega}= (n+1) [V+\bX^{T}\bX]^{-1}$. For estimating $\mcG$ from this non-sparse distribution, we generate $2,000$ independent  samples from the posterior and 	consider $(\psi^L_{ab},\psi^U_{ab})$, the 50\% credible interval for each $\psi_{ab}$. If $\psi^L_{ab}<0<\psi^U_{ab}$, then we do not include edge $(a,b)$ in $\hat{\mcG}$; if the interval excludes zero, we include the edge.
		
		\subsection{Graphical Horseshoe (GHS)}\label{ssc: graphical Horseshoe}
		The Graphical Horseshoe (GHS) model by \cite{li2019graphical} extends the GL shrinkage framework to the modeling of the precision matrix by considering a HS prior on the off-diagonal elements of $\Oomega$ and an uninformative prior on the diagonal elements. The prior is  specified element-wise such that $p(\omega_{ii}) \propto 1$ and $\omega_{ij} \sim N(0,\lambda_{ij}^2\tau^2 )$ ($i<j$), with the symmetry restriction forcing  $\omega_{ij}=\omega_{ji}$. Following the standard HS model, the local and global shrinkage parameters come from $ \lambda_{ij} \sim C^+(0,1)$ and $\tau \sim C^+(0,1)$.	Importantly, \cite{li2019graphical} show that this prior is proper despite the flat prior on the diagonal elements due to the role of the positive definite constraint. The authors have published MATLAB code (\url{https://github.com/liyf1988/GHS}) that we translated  to R  for our use. Their  Gibbs sampling algorithm updates each column of $\Oomega$ conditionally on the others, precluding parallelization and  resulting in longer run times for large $p$. We implement the GHS model by obtaining $2,000$ posterior MCMC samples after $500$ burn-in iteration. The estimated partial correlation matrix $\hat{\PPsi}$ is obtained by rescaling the posterior mean of $\Oomega$. As neither  $\hat{\Oomega}$ nor the sampled values $\Oomega$ contain exact zeros, we estimate the graph $\hat{\mcG}$ using the $50\%$ credible interval as recommended by \cite{li2019graphical} and described in Section \ref{ssc: Inverse Wishart}.

		\subsection{G-Wishart Prior (BDGraph)}\label{ssc: BDG}
		
		\cite{dawid1993hyper} introduced the hyper inverse Wishart (HIW) prior as the conjugate prior distribution for $\SSigma$ with respect to the graph $\mcG$.There has since been an abundance of articles regarding the use of this prior in GGM \citep[e.g.,][]{roverato2002hyper,carvalho2007simulation,lenkoski2011computational,wang2012efficient,banerjee2015bayesian}. Unlike GHS which imposes sparsity in $\Oomega$ through shrinkage, the G-Wishart prior jointly models the graph $\mcG$ and the precision matrix $\Oomega$, so that $\Oomega$ has the zero pattern consistent with the graph. Hence, given a graph $\mcG$, the precision matrix $\Oomega$ is constrained to the cone $\vmathbb{P}_{\mcG}$ of symmetric positive definite matrices with elements $\omega_{ij}$ equal to zero for all $(i, j)\notin E$. The G-Wishart distribution $W_{\mcG}(b, D)$ represents the conjugate prior distribution for $\Oomega$ that imposes the conditional independence relationships encoded in $\mcG$; $b$ and $D$ represent the prior degrees of freedom and the prior scale matrix similar to the standard Wishart distribution. The general sampling strategy for this involves sampling for $\mcG$ marginally over $\Oomega$ \citep{carvalho2007simulation}. We implement the continuous time birth-death MCMC algorithm from the package BDgraph \citep{mohammadi2019Rpackage}.	We  consider $20,000$ iterations from the sampling algorithm with $8,000$ burn-in iterations, and we use a non-informative prior on each edge in the graph. All other model settings  and sampling choices are set to the defaults specified in the package. An estimate for $\PPsi$ is obtained by re-scaling the estimated $\Oomega$, and $\hat{\mcG}$ comes for the median graph based on the estimated posterior link inclusion probabilities $P((a,b)\in E)$.

		\subsection{The Graphical Lasso (GLASSO) }\label{ssc: graphical lasso}  
		In addition to the previously discussed Bayesian approaches, we explore two commonly used frequentist approaches. The first is the Graphical Lasso \citep[GLASSO;][]{friedman2008sparse} that uses  $\ell_1$-regularization to control the number of zeros in the precision matrix. Specifically, GLASSO minimizes the objective function $\log |\Oomega | - \text{tr}(\bS \Oomega) - \varrho {\|\Oomega \|}_{\ell_1}$, over non-negative definite matrices $\Oomega=\SSigma^{-1}$, where $\bS=n^{-1}\sum _{i=1}^{n}( \bX_i - \bar{\bX})(\bX_i -\bar{\bX})^{T}$ is the empirical covariance matrix and $\varrho$ acts as the penalty parameter. We implement GLASSO using the package CVglasso \citep{GallowayCVglasso} under the default choices. The package uses cross validation to select a tuning parameter value for $\varrho$.	We then re-scale the estimated $\hat{\Oomega}$ to get the partial correlation matrix $\hat{\PPsi}$. The graph estimate $\hat{\mcG}$ is determined by the sparsity pattern of $\hat{\Oomega}$.

		\subsection{The Meinhausen Approach with Lasso (LASSO)}\label{ssc: meinhausen lasso}
		As noted previously, \cite{meinshausen2006high} performed parallel regressions using the LASSO penalty to estimate the neighborhood of each node. For the $a^{th} $ node-wise regression, the LASSO estimate is  $\hat{\bbeta}_{a} = \argmin_{\bbeta} \left\{n^{-1} \|X_a-\bX_{B}\bbeta\|_2^2+\varrho^{(a)}\|\bbeta\|_{\ell_1}\right\}$ where $\varrho^{(a)}$ is a penalty parameter selected through cross validation.
		The neighborhood estimate is defined by the non-zero coefficient estimates, $\hat{\ggamma}_a = \{b\in V : \hat{\bbeta}_{a}^{(b)} \neq 0 \}$. After obtaining $\hat{\bbeta}_{a}$ and $\hat{\ggamma}_{a}$ for all nodes, $\mcG$ and $\PPsi$ are estimated as in Section~\ref{ssc:postproc 1}.

		\section{Simulation Study}\label{sc: numericals}
		
		We performed numerous simulation studies to compare our proposed SUBHO approach to the previously discussed competitors. The  comparison metrics, to be discussed next, relate to graph  sparsity structures  and the partial correlations. Analyses for SUBHO, BASAD, GHS, and Inverse Wishart rely on R code that we developed, while	Hyper-$g$ (BAS), BDGraph, GLASSO, and LASSO rely primarily on existing R packages.
		Our Projpred-HS implementation performs  the analysis using an adjusted version of the author's Github code to avoid the library dependency errors, and includes  HS sampling within STAN.

		\subsection{Simulation Design}\label{ssc:simulation}

\begin{figure}
			\centering
			\includegraphics[width=\textwidth]{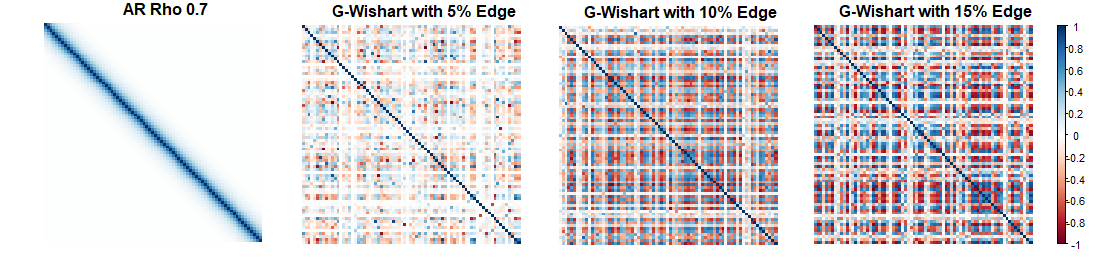}
			\caption{True Covariance Structures in $(n,p)=(75,75)$ Case}
			\label{fig:cov_structures_used}
		\end{figure}

		We use two separate covariance structures to simulate the data from  multivariate normal distributions. For the first choice we have an  auto-regressive structure with $\rho = 0.7$. In this AR(1) case the resulting graph structure will be non-zero only for the first off-diagonal elements. The second choice is to generate the true $\Oomega$ using the G-Wishart distribution, conditionally on a randomly generated graph. We generate $\mcG$ by sampling edge indicators from Bernoulli with $p_{\mcG} = \{0.05, 0.10, 0.15\}$. Given $\mcG$, we draw $\Oomega\sim W_{\mcG}(b,D)$, where we set the degrees of freedom $b$ to be $3$ and the scale to be the identity. We note that a diagonal prior scale matrix pushes the non-diagonal elements of $\Oomega$ (and consequently, $\PPsi$) towards small values, which translates to small effects and challenging estimation across all methods. Finally, we re-scale the $\Oomega$ such that $\SSigma=\Oomega^{-1}$ has unit diagonal. We illustrate our generated covariance structures for $n=p=75$ in Figure~\ref{fig:cov_structures_used}. We consider $4$  choices for the number of samples ($n$) and the number of nodes/variables ($p$):  $(n,p) \in \{(75,75),(150,75),(75,150),(150,150)\}$. For each $(n,p)$ and each covariance choice, we generate $10$ datasets from the corresponding mean zero multivariate normal distribution. Assessing the performances of different methods on graph estimation, we use a collection of accuracy measures related to the accuracy of the partial correlation estimation and the graph recovery. For each method, all performance measures are calculated based on an average of $10$ replicated datasets under each setting.

		\subsection{Graph Estimation Recovery Comparison by FDR }\label{sssc:FDRstuff}
		
		\begin{figure}
			\centering
			\includegraphics[width=\textwidth]{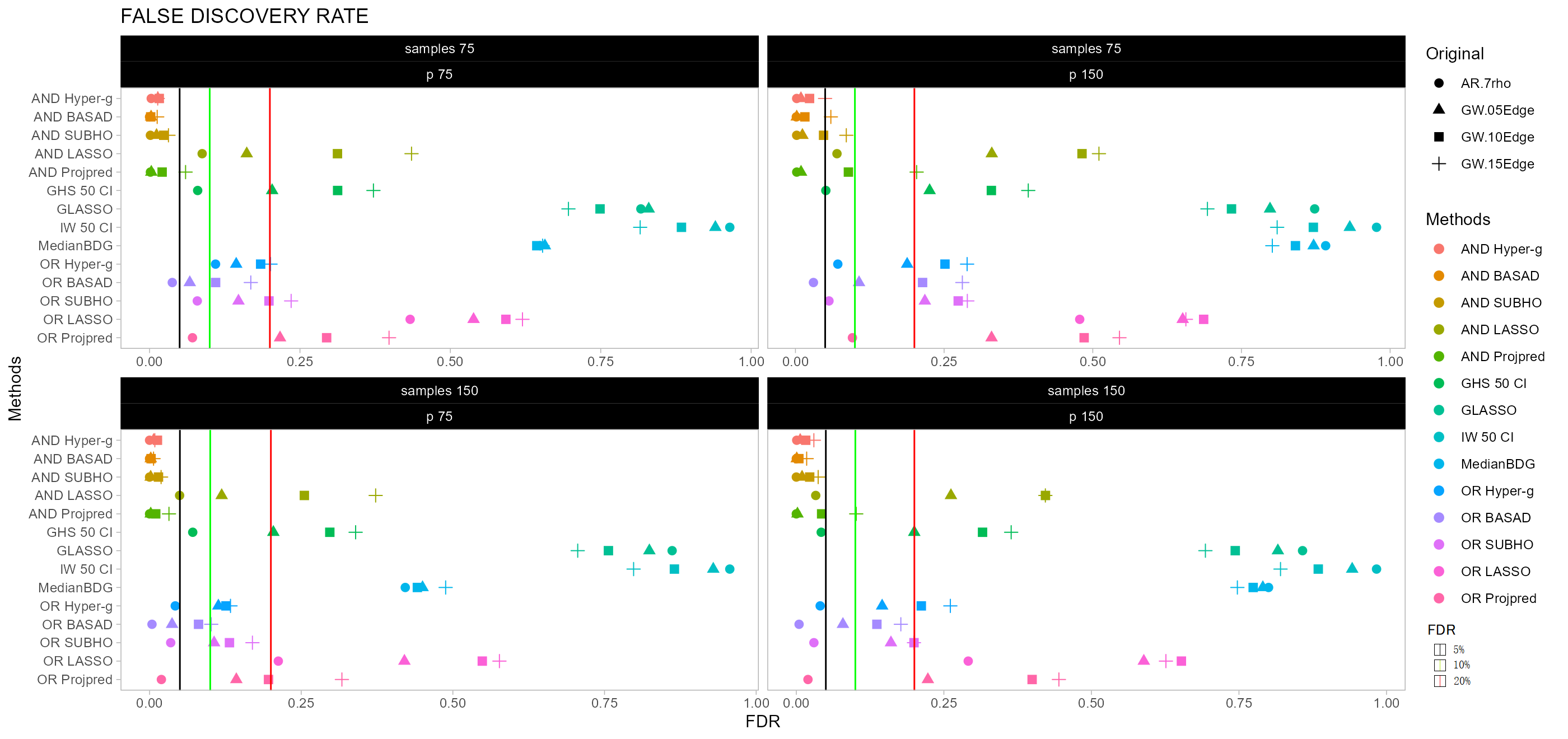} 
			\caption{False Discovery Rate Comparison}
			\label{fig:FDR}
		\end{figure}
		
		First, we use the false discovery rate (FDR) as a tool of assessing the accuracy of the graph estimates $\hat{\mcG}$. FDR is the expected proportion of discoveries (rejected null hypotheses) that are incorrect rejections of the null \citep{benjamini1995controlling}. Equivalently, the FDR is the expected ratio of the number of false positive edges  to the total number of true positive edges. Results are shown in Figure~\ref{fig:FDR}.

		FDR comparisons show that the Hyper-$g$ (BAS) AND, BASAD AND, SUBHO AND, and  Projpred-HS AND control FDR at less than $0.1$ for all $16$ scenarios ($4$ data types and $4$ $(n, p)$ combinations). Here, the AND attached to a method indicates that these are graph estimates constructed by combining the neighborhoods of each node using the AND rule $\hat{E}^\land$ as in Equation~\eqref{eq: symmetrization}. An exception to above FDR control rates is that Projpred-HS AND has an FDR around 10\% for  G-Wishart$(p_\mcG = 0.15)$ when $(n,p) =(150, 150)$ and more than 10\% for G-Wishart$(p_\mcG = 0.15)$ with $(n,p) =( 75, 150)$. LASSO AND is only able to control the FDR when data are drawn from an AR structure. Similarly, the denser estimates from the OR approaches fail to control FDR, except for under the simplest AR choice. Among methods that are based on fully estimating $\Oomega$ and/or $\mcG$ structure (GLASSO, BDG, IW and GHS), all yield substantially denser graphs than the AND methods with substantially higher FDR, often in excess of $50\%$. 
		
		Based on Figure~\ref{fig:FDR}, we conclude that the fully Bayesian methods (GHS, BDG, IW) and the typical frequentist approaches (GLASSO and AND/OR LASSO) fail to control the FDR, thus can not be trusted. Bayesian parallel regression approaches that combine neighborhoods with the AND operator perform the best. For the rest of this article, 
		we no longer consider the graph estimations created by combining neighborhoods through the OR relationship, and 
		we focus only on  AND versions of these methods due to their superior FDR performance. In general, there is not much difference in FDR between AND estimates from Hyper-$g$, BASAD, and SUBHO. In quite a few cases under the most dense $(p_\mcG = 0.15 )$ scenario, Projpred-HS AND yielded higher FDR than the other parallel Bayesian methods.

		\subsection{Graph Estimation Recovery: Comparison by TPR }\label{sssc:TPRstuff}
		We now consider the  True Positive Rate (TPR) as a metric for comparing  accuracy. In our case, 	TPR refers to the probability of  estimating the edge in $\hat{\mcG}$, conditionally on it truly being an edge in the underlying  ${\mcG}$. 
		TPR is equivalent to the power to detect an edge. As noted previously, the OR estimates had substantially higher FDR than their AND counterparts and are  excluded  from these comparisons.

		\begin{figure}
			\centering
			\includegraphics[width=\textwidth]{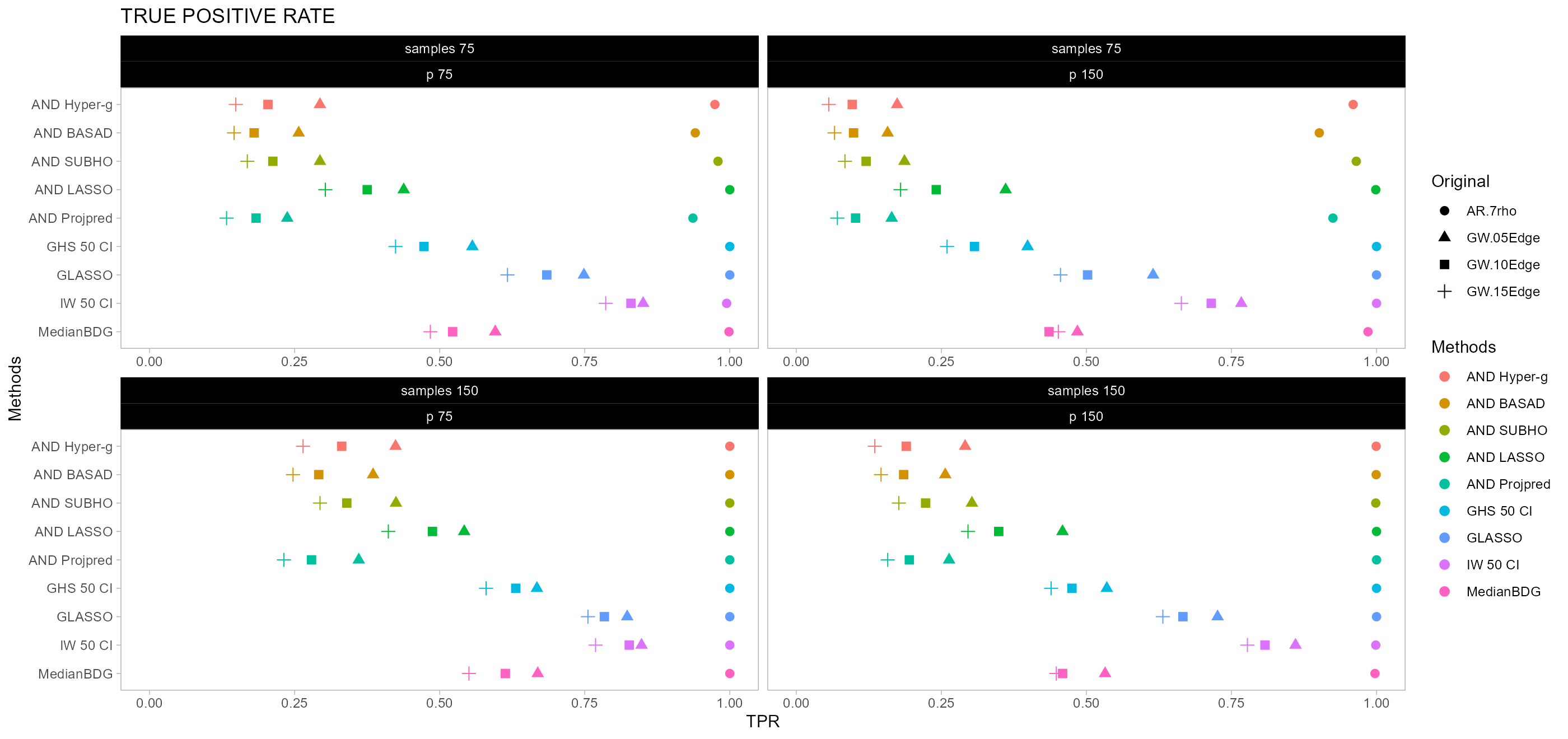} \caption{True Positive Rate Comparison} 
			\label{fig: TPR}
		\end{figure}

		From Figure~\ref{fig: TPR} we observe that across all $(n,p)$, all methods are performing good in terms of power/TPR for the AR covariance. Obviously, this scenario offers the least challenges with few  true edges and all having high partial correlations ($\rho = 0.7$).  GLASSO has the highest power followed by the BDG, IW and GHS. Under the  most challenging scnenario with G-Wishart$(p_\mcG = 0.15)$,  these methods achieve higher TPR than the corresponding Nbd-Sel approaches. However, we need to be cautious in our interpretations as these results are accompanied by  high FDR. Exploring the node-wise regression methods, LASSO AND, which also failed to control FDR, has the highest power. Among the four Bayesian methods with their AND estimators, Projpred-HS tends to have the lowest power for the most dense graph G-Wishart$(p_\mcG = 0.15)$. Hyper-$g$ (BAS) and SUBHO tend to have higher power  across the majority of cases. In view of the  FDR and TPR results, we conclude that the proposed SUBHO AND has the most optimal performance for estimating  the graph structure $\mcG$ in the sense that it typically has the highest power among methods that control the FDR at a reasonable rate.

		\subsection{Partial Correlation Estimation Accuracy}\label{sssc: Partial Correlation}
		
    	To obtain the estimation error of the partial correlations, we divide the problem into two parts. First, we view the estimation error for the positions where we had true zeros in $\PPsi$ as $MSE_{\{zero\}} = \sum_{a<b} (\hat{\psi}_{ab}-{\psi}_{ab})^2 \times \vmathbb{1}({\psi}_{ab} =0 )$. Second, we consider  the error for the positions with edges (non-zero elements) as $MSE_{\{\neq zero\}} = \sum_{a<b} (\hat{\psi}_{ab}-{\psi}_{ab})^2 \times \vmathbb{1}({\psi}_{ab} \neq 0 )$. Finally, we combine these together to get the total mean squared error ($MSE$) for the estimation of $\PPsi$ as $MSE_{Total} = \sum_{a<b} (\hat{\psi}_{ab}-{\psi}_{ab})^2 = MSE_{\{zero\}}  + MSE_{\{\neq zero\}}$.
		
		\begin{figure}
			\includegraphics[width=\textwidth]{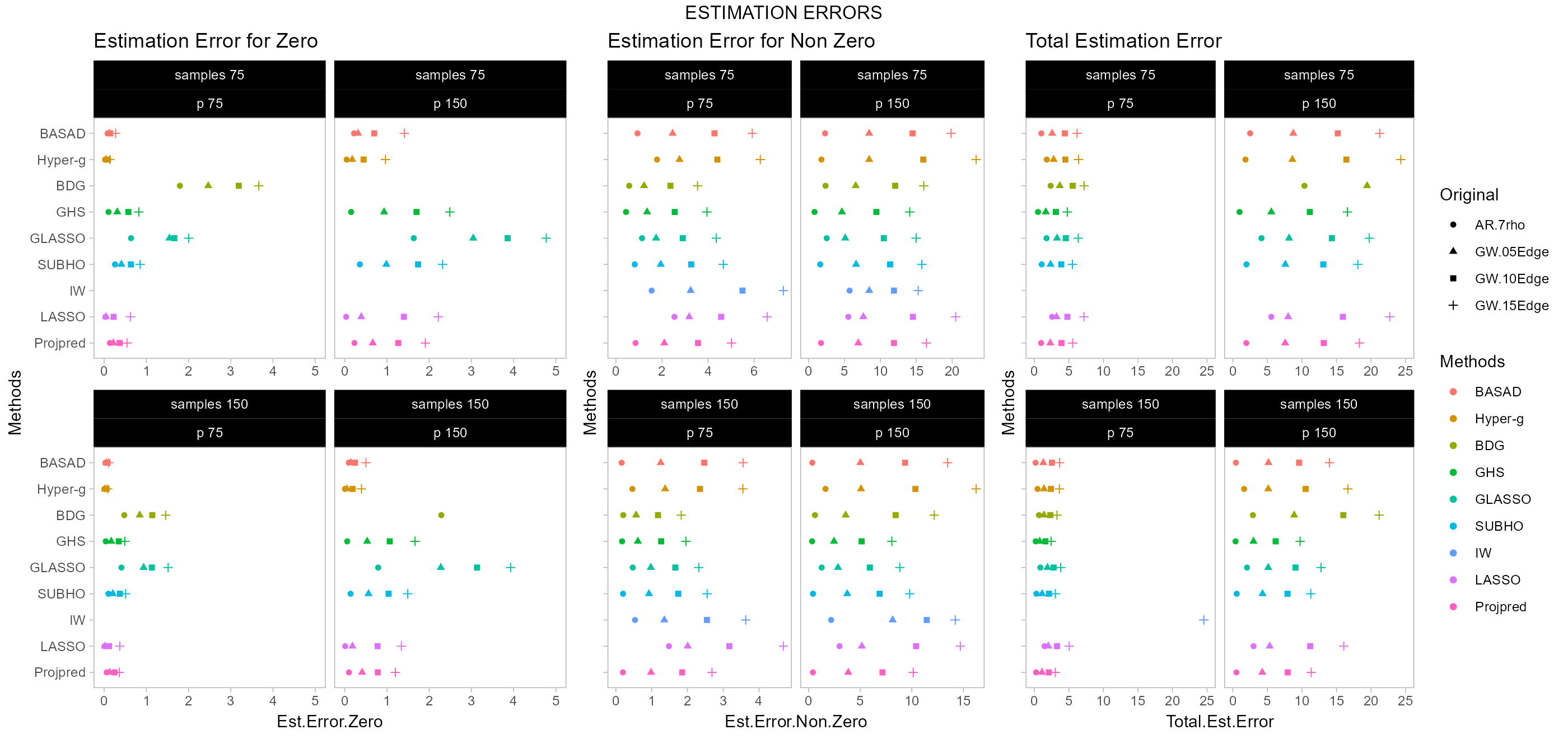} \caption{Comparison of Estimation Error for Partial Correlation Matrices}
			\label{fig: MSE Zero and non Zero}
		\end{figure}

		We  first consider $MSE_{\{zero\}}$. From the left third of Figure~\ref{fig: MSE Zero and non Zero}, one can clearly see that the selection prior methods, BASAD and Hyper-$g$ (BAS), yield the best estimates of the zero partial correlations. LASSO and GHS, as well as  SUBHO and Projpred (which use very similar HS choices to obtain $\hat{\bbeta}_a$), perform well and similarly to each other. As in graph estimation, the $\PPsi$ estimates  from BDG and GLASSO are much denser and come with substantial errors for the zero correlations. Estimation errors for IW  and BDG are too large to fit into the figure margin for $p=150$. For the non-zero partial correlations, most methods perform  similarly in terms of $MSE_{\{\neq zero\}}$. The sparsest BASAD and Hyper-$g$ (BAS) have somewhat elevated error relative to the BDG, GHS, and the HS  methods (SUBHO and Projpred). Finally, in terms of  $MSE_{Total}$, we see that GHS produces the best partial correlation recovery. In general, this is what might be expected from a fully Bayesian approach designed to accurately estimate the concentration matrix by implementing variable shrinkage. However, as we have shown previously, it does not perform well for graph estimation. The HS estimated partial correlations are the next best set of estimates (typically, they are very close in performance to GHS), and we have shown that these results can simultaneously be used to gain insight on the graph while controlling the FDR.

		\subsection{Conclusions from Simulations}\label{ssc:Discussion}
		
		In Appendix A of the Supplementary Materials, we include some further details about computational choices and the associated computational times for each method. We make a few general comments here. Among the Nbd-Sel methods, LASSO is much faster than the corresponding Bayesian methods, but among the Bayesian choices SUBHO is the fastest. Projpred is multiple orders of magnitude slower due to the LOOCV calculation required for each model accuracy $u(\mcM)$. The fully Bayesian approach BDG and GHS do not facilitate parallel computations and are thus slower than SUBHO.
		
		We close this simulation comparison with a few final comments. Overall, most of the methods fail to control FDR at any reasonable rate. In particular, GLASSO,  which is a commonly used graph estimation method, consistently yields overly dense covariance matrices. GHS produces the best estimates of the partial correlations. However, its credible interval strategy yields overly dense graphs, and it is also somewhat slower than the Nbd-Sel approaches. All OR methods fail to control FDR.  
		On the other hand, our proposed methods combining node-wise regressions with Bayesian regression  and the AND  thresholding controlled FDR at $0.05$ in the majority of the scenarios and at $0.10$ in all cases under consideration, but this comes at a cost of reduced TPR. Among these Bayesian parallel regression approaches, we believe SUBHO AND performs the best in terms of controlling FDR and attaining higher power.

		\section{Application to METABRIC-TNBC Data}\label{sc: TNBC data analysis}
		
		In this section, we estimate a gene network from a breast cancer data set. 
		Breast cancer is one of the most common  cancers effecting women. 
		Triple-negative breast cancer (TNBC) refers to breast cancer (BC) cells that do not have estrogen receptor (ER) or progesterone receptors (PR) and  also do not produce the protein HER2 (human epidermal growth factor receptor 2). Many common hormone therapy and drugs that target HER2 are ineffective for  treating TNBC. Recently, researchers have been studying biological characteristics such as pathology, histology and  survival rates for TNBC \citep{jhan2017triple, li2017triple}. However finding the optimal treatment strategy for patients with TNBC still remains an active research area \citep{bianchini2022treatment}. The data we use for this investigation come from  BRCA-METABRIC \citep{curtis2012genomic,pereira2016somatic,rueda2019dynamics}.
		
		The data consists of continuous, log-transformed mRNA $Z$-Scores (standardization was done based on all samples using Illumina Human v3 microarray) for $1,906$ subjects and $24,368$ genes. Removing missing values leaves $1,894$ BC subjects, and we focus on the  $n=299$ triple negative patients (ER$-$, PR$-$ and HER2$-$). Of the original $24,368$ genes, we select those genes that are differentially expressed by TNBC status. We fit a sequence of one-predictor logistic regression models with the $Z$-score for each gene as the predictor and TNBC status as the outcome (comparing the $n=299$ TNBC patients versus the remaining $n=1569$ patients without TNBC as controls for variable selection). We selected the $p=1,000$ most differently expressed genes by ranking the $p$-values in descending order.  We construct the gene expression network by applying all methods considered previously to this data with $n=299$ patients and $p=1000$ genes. However, the Projpred method was too slow to run in this example. To get a sense of the time required for this method, we ran $10$ of the necessary $1000$ nodal regressions, but each took approximately $23$ hours to produce the $\hat{\bbeta}_a$ and $\hat{\ggamma}_a$ estimates.

		\subsection{Network Exploration}\label{sssec: network exploration}

		\begin{figure}
			\centering
			\includegraphics[width=\textwidth]{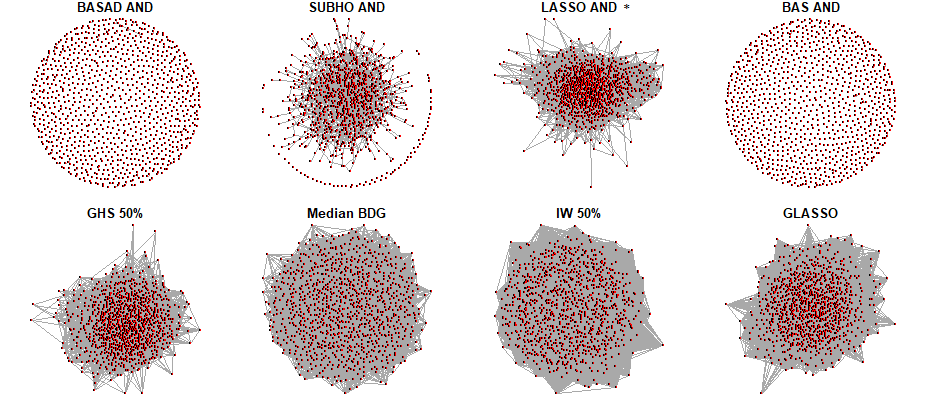}   
			\caption{Visual Representation of estimated METABRIC TNBC Gene Networks.\\ *Removed $6$ singleton nodes from LASSO AND network for appropriate figure scaling.}
			\label{fig:TNBCdata_gene_network 1}
		\end{figure}

		Figure~\ref{fig:TNBCdata_gene_network 1} is a graphical representation of the estimated networks from each method. With $p=1000$, there are $499,500$ possible edges that could be included in the graphs, but we  anticipate that underlying network should be fairly sparse. The graphs for BASAD AND ($210$ edges estimated) and Hyper-$g$ (BAS) AND ($145$ edges) include  very few connections, suggesting a lack of statistical association across most genes. The graphs from GHS $(11,629)$, BDGraph $(13,356)$, IW $(216,771)$ and GLASSO $(66,696)$, in contrast, are quite dense. These four networks suggest that most nodes/genes share connections to multiple other gene. The LASSO AND $(10,807)$ finds a graph with a similar number of edges to GHS and BDG, but the network structure is somewhat different due to the presence of six singleton nodes that do not interact any others nodes. The graph from SUBHO AND indicates that some of these genes are not associated with  the rest. The outer periphery in the figure represents those genes while  inner cluster shows the ones that form a connection. SUBHO AND finds a manageable $1,586$  number of estimated edges.

		\subsection{Most Connected Genes}\label{ssc: top genes tnbc}
		
\begin{figure}
			\centering
			\includegraphics[width=\textwidth]{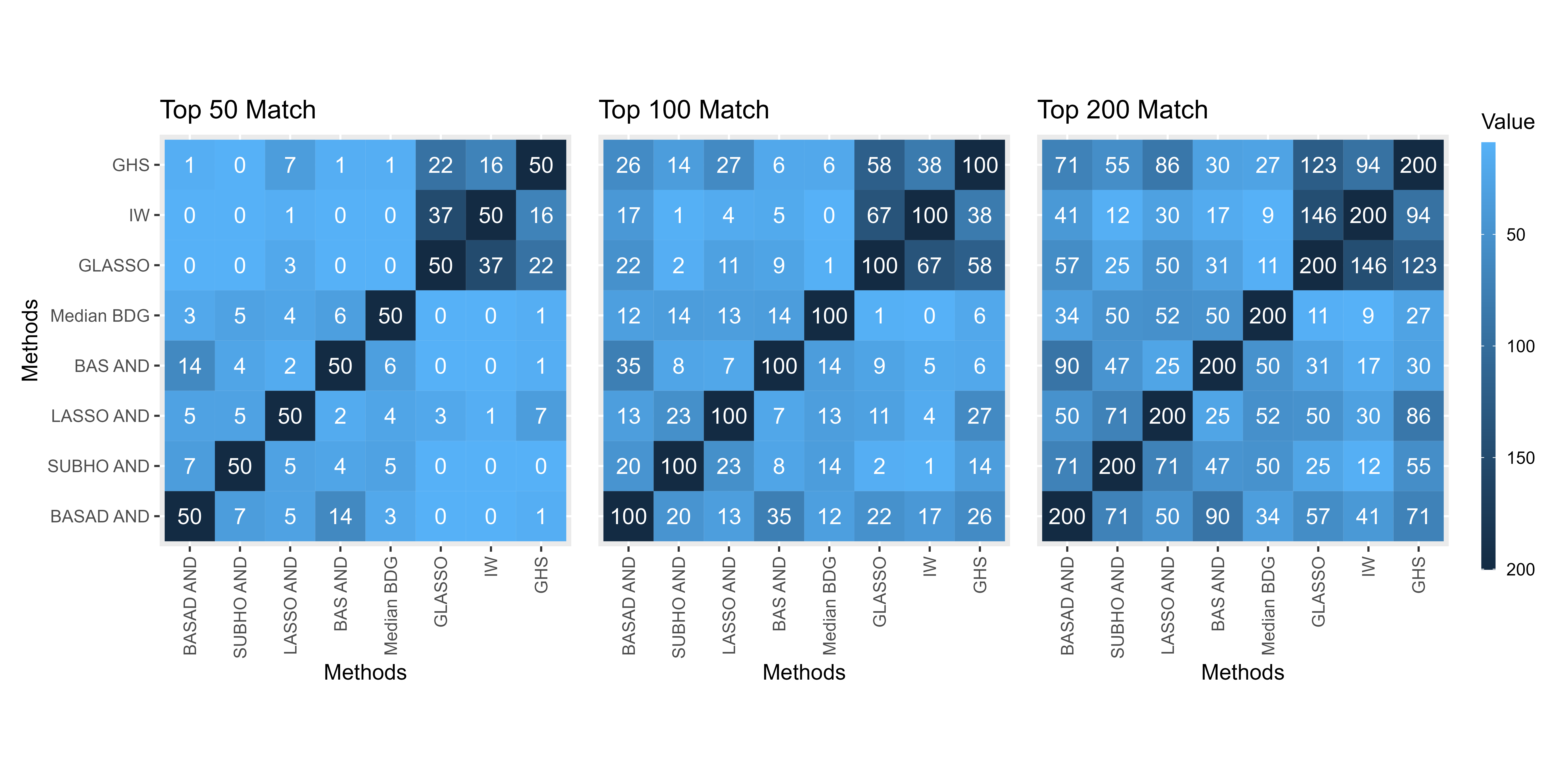}
			\caption{Comparison of Gene Name Match}
			\label{fig:Gene name mis-match TNBC}
		\end{figure}

		Based on these estimated networks, we seek to obtain a set of the most connected genes for each method. We consider the number of connections to gene $a$ as given by $G_{a}=\sum_{b=1}^{p} \hat{\mcG}_{ab}$. For the $a^{th}$ gene, a higher value of $G_{a}$ would imply higher connectivity to the rest of the genes. By considering the $K$ highest values of $G_{a}$, we obtain the top $K$ most connected genes. To break any ties among genes with the same  $G_{a}$ score, we use $P_{a}=\sum_{b=1}^{p} | \hat{\psi}_{ab} |$ to  measure the sum of all absolute partial correlations. Obtaining the top $K\in\{50,100,200\}$ most connected genes from each method, we consider   the overlap in these most connected genes across all pairs of methods. These  similarities  are shown in Figure~\ref{fig:Gene name mis-match TNBC}. Interestingly GLASSO, IW and GHS tend to produce sets of most connected genes  that are  more similar to each other than the other approaches. This may not be surprising since  these three methods are  somewhat similar since their estimation targets are the precision matrix $\Oomega$. We note that majority of methods are finding highly disparate  sets of top $K$ genes, indicating substantial differences in the inferences from each analysis.
		
		\subsection{Gene Set Enrichment}\label{ssc:GSE TNBC}

		In light of the substantial differences between the genes identified as most connected across methods, we use the  functional mapping and annotation (FUMA) approach for summarizing GWAS (genome wide association studies) studies by \cite{watanabe2017functional} to investigate the biological relevance of the estimated gene sets. FUMA integrates data from multiple bio-repositories, and genes are prioritized via positional mapping based on annotations obtained from ANNOVAR. To test for over-representation of biological functions based on gene annotations, FUMA screens hallmark gene sets from MSigDB. \cite{liberzon2015molecular} introduced hallmark gene sets that summarize and represent specific well-defined biological states or processes and display coherent expression; they also discuss how hallmarks provide a better portrayal of biological space for gene set enrichment (GSE) analysis while reducing noise and redundant information. Hallmark gene sets are linked to their corresponding founder sets and can be explored for deeper investigation. Each hallmark pathway has a list of associated genes that can be obtained from either MSigDB or KEGG database. Our analysis includes enrichment with hallmarks for finding significant overlap with these gene sets. For each of the network methods, we submitted the list of the $K=100$ most connected genes for testing against the GWAS mapped genes (a total of $57,241$ protein-coding reference genes) using hypergeometric enrichment tests. FUMA accounts for multiple testing in this step, and gene sets with an FDR adjusted $p$-value less than $0.05$ were considered significant evidence of enrichment. 
		
\begin{figure}
			\centering
			\includegraphics[width=11.5cm]{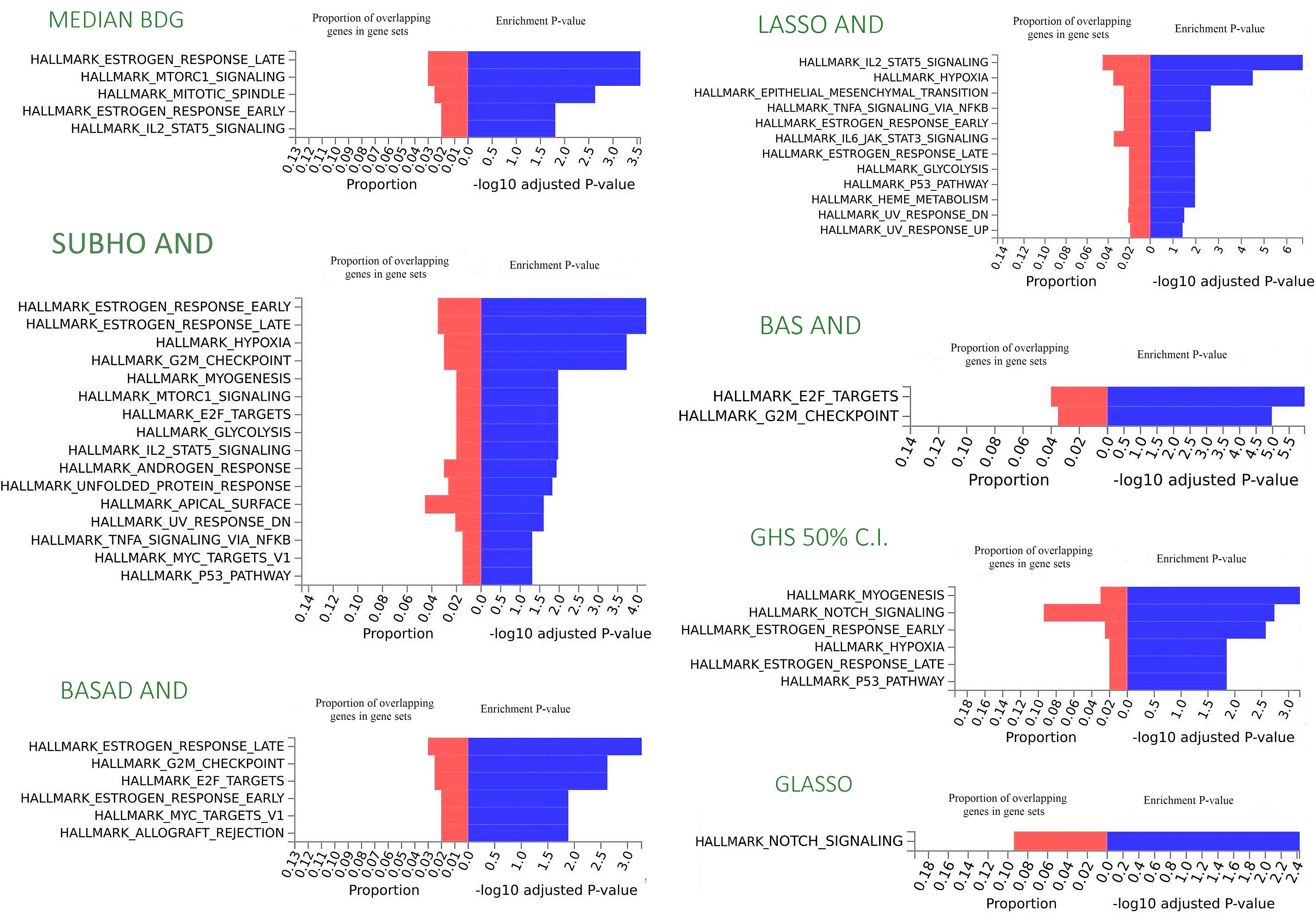}
			\caption{Enrichment results with significant hallmarks with top $100$ TNBC genes.} \label{fig:TNBCdata_hallmark_gene_sets}
		\end{figure}

		GSE results are shown in  Figure~\ref{fig:TNBCdata_hallmark_gene_sets}. SUBHO AND is associated with $16$ significantly enriched hallmark gene sets which is more than any of the other methods. We take this as providing biologically-based validity to the results from our method. LASSO identifies $12$ significant hallmark gene sets, while BASAD and GHS both find $6$; all other methods identify five or fewer. We also see that the majority of hallmark gene sets identified by competing methods are also discovered by SUBHO.
		
		Based on Table~$1$ of \cite{liberzon2015molecular}, we further investigate the process category for those hallmark gene sets that were identified (Figure~\ref{fig:TNBCdata_hallmark_gene_sets}) from the top $100$ most connected genes from the SUBHO AND network. Many of these hallmark sets are signaling processes, including Estrogen Response Early and Late, MTORC1, IL2 STAT5, TNFA Signaling via NFKB, and Androgen Response. Hypoxia and Unfolded protein response are classified as in the process category of pathways. G2M Checkpoint, P53 Pathway, E2F Targets, and MYC Targets V1 are related to cell proliferation. Apical Surface corresponds to the cellular component which is related to membrane proteins in the apical domain. Myogenesis is related to muscle differentiation, and Glycolysis is  related to metabolic process. In the Appendix, we list and further investigate the individual genes (a total of $34$) that are responsible for the enrichment of these $16$ hallmark gene sets. In particular, we note that all of these genes have been reported in the literature as related to breast cancer and/or TNBC. Again, we believe this further supports the biological plausibility of the estimated network from SUBHO AND. The Appendix contains additional results related to this enrichment analysis as well as an analysis of a second gene expression data set.

		\section{Discussion}
		
		In this project we have approached the problem of estimating  Gaussian graphical models through the regression models associated with each node.  Under this Nbd-Sel framework, the  available parallelization leads to efficient Bayesian computation as each regression model can be fit quickly.  We have shown this using both continuous shrinkage priors and selection priors (BASAD and hyper-$g$).  While the graph can be easily estimated under the selection choices, it is non-trivial to construct $\hat{\mcG}$ under a horseshoe prior for $\bbeta_a$.  Unlike the time-consuming prediction-based strategy of \cite{williams2018bayesian}, we propose in SUBHO a simple model space reduction by considering the ranks of the partial regression coefficients.  This leads to a dramatically faster result with no appreciable decrease (and potentially a small increase) in graph estimation accuracy.

		One of the primary contributions in this work is the use of the step-up model selection.  Using the regularized estimates of the partial regression coefficients from the HS model fit, each predictor is ranked, and the only models considered are those constructed by taking the $k$ highest ranked predictors. This intuitive choice dramatically reduces the $\ggamma_a$ model space considered from $2^{p-1}$ models down to $p-1$ models (or often $K<p-1$). Slight alterations to this strategy could be considered to allow further investigation within the model space. For instance, in addition to the $k$ predictor model given by the predictors $\{R(1),\ldots,R(k)\}$, one could also consider models such as $\{R(1),\ldots,R(k-1),R(k+j)\}$, $j=1,\ldots,J$ for small $J$, yielding a model space of size $(p-1)(J+1)$. Many similar choices based on the HS-estimated ranks could be considered to add further flexibility in the model space searched, while maintaining computational scalability.

		An important conclusion from this investigation is that  in this context the methods considered frequently produce very different results on the same data set.  As shown in the TBNC analysis, GLASSO and the  Bayesian methods that are designed to model the precision matrix estimate graphs that are substantially more dense than the Nbd-Sel approaches with over 10,000 edges on $p=1000$ genes.  This is consistent with the simulation results where these methods detect substantially more edges than the Nbd-Sel strategy. But these simulation results also suggest that a large portion of these detected edges are likely to be false positives.  While the Bayesian node-wise regression choices when combined with the AND rule for graph construction are able to maintain FDR at reasonable levels, the resulting power (TPR) and total number of edges detected are substantially lower.  Additional investigation into this effect and how one might further adjust the SUBHO strategy to increase power is needed.

		In many cases one is primarily interested in inference on the graph structure $\mcG$.  However, we additionally focus on concurrent modeling of the partial correlations $\PPsi$ to get a deeper sense of the dependence across nodes.  The estimated graph $\hat{\mcG}$ only indicates whether or not variables are conditionally related but does not convey information about the magnitude or direction of this relationship.  We highlight that these partial correlations can be directly estimated from the regression models fit for Nbd-Sel and investigate the estimation accuracy.  Somewhat surprisingly, using the parallel regression approach estimates $\PPsi$ almost as accurately as GHS, a state-of-the-art strategy for modeling sparse precision matrices.  This suggests that $\hat{\PPsi}$ can also be used in tandem with $\hat{\mcG}$ to draw important inferences related to the signs and strengths of the identified edges.

		\appendix
		
		\section{Computational Time}\label{appendix: time taken}
		In this section we discuss the computation time used for each method. Here, we have tabulated the time in minutes that each of the methods has taken to provide the estimated $\mcG$ and $\PPsi$.
		We have run the SUBHO, LASSO, BASAD, Hyper-$g$ (BAS), and Projpred-HS in parallel using R. For these Nbd-Sel approaches, individual regressions were run in parallel exploiting the conditional independence across graphical neighborhoods.  All computations are run on a Intel(R) Core(TM) $i7-8700$ CPU $@3.20$GHz with $32$GB RAM, and those that make use of parallel regressions are run using eleven out of the twelve nodes. 
		
		Let us note first that by parallel, we primarily refer to parallelization across graphical neighborhoods. To that end, the Nbd-Sel methods are able to simultaneously fit up to eleven of the ${X}_a = \bX_{B}\bbeta_a + \epsilon_a$ regression models at a time. In contrast, GHS is a fully Bayesian approach that relies on a Gibbs sampler to update a block of parameters, conditionally on the current value of the remaining parameters, in cycle. Consequently, this can not be parallelized across graphical neighborhoods in the same way. Similarly, BDGraph uses an efficient MCMC algorithm and does not have node-wise regressions that can be  parallelized over computational nodes, thus we employed only one node of the CPU for computation.
		For Inverse Wishart, the times reported are basically the time required to generate $2000$ samples form the distribution. GLASSO (CVglasso) considers estimation of the entire $\Oomega$ in a single step and is not intended to run in parallel across graphical neighborhoods. We note that the cross validation can be run in parallel, but our implementation did not include this.

		\begin{table}
			\tiny
			\centering
			\caption{The summary of computational time of all methods}
			\label{Computational Time}
			\begin{tabular}{*{8}{l}}
				\hline
				\hline \\
				\multicolumn{1}{p{1.5cm}}{True Cov Str.} & 
				\multicolumn{1}{p{.5cm}}{$n$} &
				\multicolumn{1}{p{.5cm}}{$p$} &
				\multicolumn{1}{p{1cm}}{SUBHO} &
				\multicolumn{1}{p{1cm}}{BASAD} &
				\multicolumn{1}{p{1cm}}{Hyper-$g$ (BAS)}  & 
				\multicolumn{1}{p{1cm}}{Projpred-HS} &
				\multicolumn{1}{p{1cm}}{LASSO} \\
				\hline \\
				\multicolumn{1}{p{1cm}}{Languages} & 
				\multicolumn{1}{p{.5cm}}{} &
				\multicolumn{1}{p{.5cm}}{} &
				\multicolumn{1}{p{1cm}}{R} &
				\multicolumn{1}{p{1cm}}{R} &
				\multicolumn{1}{p{1cm}}{C \& R} & 
				\multicolumn{1}{p{1cm}}{Stan \& R} &
				\multicolumn{1}{p{1cm}}{C$++$ \& R} \\
				\hline \\
				\multicolumn{1}{p{1cm}}{Parallel} & 
				\multicolumn{1}{p{.5cm}}{} &
				\multicolumn{1}{p{.5cm}}{} &
				\multicolumn{1}{p{1cm}}{\checkmark} &
				\multicolumn{1}{p{1cm}}{\checkmark} &
				\multicolumn{1}{p{1cm}}{\checkmark} & 
				\multicolumn{1}{p{1cm}}{\checkmark} &
				\multicolumn{1}{p{1cm}}{\checkmark}\\
				\hline \\
				AR $\rho=0.7$	&	75	&	75	&	0.58	&	1.78	&	6.16	&	167.66	&	0.02	\\
				GW $5 \%$ E  	&	75	&	75	&	0.59	&	1.63	&	13.36	&	299.50	&	0.02	\\
				GW $10 \%$ E	&	75	&	75	&	0.58	&	1.65	&	13.71	&	305.15	&	0.02	\\
				GW $15 \%$ E	&	75	&	75	&	0.59	&	1.64	&	13.58	&	308.73	&	0.02	    \\
				AR $\rho=0.7$	&	75	&	150	&	3.26	&	9.49	&	7.51	&	741.69	&	0.04	\\
				GW $5 \%$ E	   &	75	&	150	&	3.24	&	8.81	&	26.11	&	791.98	&	0.04	\\
				GW $10 \%$ E	&	75	&	150	&	3.26	&	8.87	&	25.33	&	807.63	&	0.04	\\
				GW $15 \%$ E	&	75	&	150	&	3.26	&	8.79	&	26.30	&	810.03	&	0.04	\\
				AR $\rho=0.7$	&	150	&	75	&	1.12	&	2.78	&	13.17	&	201.17	&	0.02	\\
				GW $5 \%$ E   	&	150	&	75	&	1.12	&	2.55	&	7.34	&	301.01	&	0.02	\\
				GW $10 \%$ E	&	150	&	75	&	1.11	&	2.59	&	7.55	&	302.70	&	0.03	\\
				GW $15 \%$ E	&	150	&	75	&	1.11	&	2.62	&	7.83	&	302.57	&	0.03	\\
				AR $\rho=0.7$	&	150	&	150	&	6.80	&	47.31	&	54.48	&	2493.46	&	0.09	\\
				GW $5 \%$ E	    &	150	&	150	&	6.83	&	23.11	&	36.88	&	1960.00	&	0.09	\\
				GW $10 \%$ E	&	150	&	150	&	6.91	&	21.26	&	35.61	&	2017.62	&	0.08	\\
				GW $15 \%$ E	&	150	&	150	&	6.91	&	21.29	&	34.26	&	2016.50	&	0.08	\\
				\hline \\
				\multicolumn{1}{p{1.5cm}}{True Cov Str.}
				& \multicolumn{1}{p{.5cm}}{$n$} &
				\multicolumn{1}{p{.5cm}}{$p$} &
				\multicolumn{1}{p{1cm}}{GHS} &
				\multicolumn{1}{p{1cm}}{GLASSO} &
				\multicolumn{1}{p{1cm}}{BDG}  & 
				\multicolumn{1}{p{1cm}}{IW}   \\
				\hline \\
				\multicolumn{1}{p{1cm}}{Languages} & 
				\multicolumn{1}{p{.5cm}}{} &
				\multicolumn{1}{p{.5cm}}{} &
				\multicolumn{1}{p{1cm}}{R}  & 
				\multicolumn{1}{p{1cm}}{Fortran, C \& R}  & 
				\multicolumn{1}{p{1cm}}{C, C$++$ \& R} & 
				\multicolumn{1}{p{1cm}}{R}   \\
				\hline \\
				\multicolumn{1}{p{1cm}}{Parallel} &
				\multicolumn{1}{p{.5cm}}{} &
				\multicolumn{1}{p{.5cm}}{} &    
				\multicolumn{1}{p{1cm}}{\texttimes}  & 
				\multicolumn{1}{p{1cm}}{\texttimes}  & 
				\multicolumn{1}{p{1cm}}{\texttimes} & 
				\multicolumn{1}{p{1cm}}{\texttimes}  \\
				\hline \\    
				AR $\rho=0.7$	&	75	&	75	&	1.54	&	0.03	&	2.98	&	0.14	\\
				GW $5 \%$ E  	&	75	&	75	&	1.57	&	0.04	&	3.08	&	0.12	\\
				GW $10 \%$ E	&	75	&	75	&	1.60	&	0.05	&	3.08	&	0.12	\\
				GW $15 \%$ E	&	75	&	75	&	1.62	&	0.06	&	3.10	&	0.12	\\
				AR $\rho=0.7$	&	75	&	150	&	14.44	&	0.33	&	23.69	&	0.55	\\
				GW $5 \%$ E	   &	75	&	150	&	14.73	&	0.61	&	24.05	&	0.55	\\
				GW $10 \%$ E	&	75	&	150	&	14.96	&	0.87	&	23.37	&	0.56	\\
				GW $15 \%$ E	&	75	&	150	&	15.00	&	0.93	&	23.86	&	0.56	\\
				AR $\rho=0.7$	&	150	&	75	&	1.54	&	0.02	&	2.93	&	0.14	    \\
				GW $5 \%$ E   	&	150	&	75	&	1.59	&	0.04	&	3.08	&	0.14	\\
				GW $10 \%$ E	&	150	&	75	&	1.60	&	0.05	&	2.90	&	0.12	\\
				GW $15 \%$ E	&	150	&	75	&	1.60	&	0.06	&	2.89	&	0.12	\\
				AR $\rho=0.7$	&	150	&	150	&	29.40	&	0.15	&	23.35	&	0.55	\\
				GW $5 \%$ E	    &	150	&	150	&	14.76	&	0.52	&	23.65	&	0.56	\\
				GW $10 \%$ E	&	150	&	150	&	14.13	&	0.71	&	23.66	&	0.56	\\
				GW $15 \%$ E	&	150	&	150	&	14.64	&	0.69	&	23.61	&	0.54	\\
				\hline    \hline     
			\end{tabular}
		\end{table}

		In Table~\ref{Computational Time}, we report the average computational time in minutes required to fit a single data set  in terms of the choice of $(n,p)$ and covariance structure. We observe that among the parallel Bayesian Nbd-Sel approaches, our SUBHO is the fastest. Among these node-wise regression methods, Projpred-HS is the slowest. This is due to the very costly leave-one-out step to evaluate the posterior predictive accuracy measures for each node-wise regression model. Hyper-$g$, while more efficient that Projpred, is also substantially slower than SUBHO and BASAD. The two fully Bayesian approaches, GHS and BDG are substantially slower than SUBHO. The penalized methods LASSO and GLASSO are very fast compared to the Bayesian competitors.

		We note that there are a couple of important limitations to this comparison.  We have assessed all methods using R to try to make the computational times as comparable as possible. However, as we previously noted,  the GHS authors posted MATLAB code for their method, which may be expected to be meaningfully faster than time required for the corresponding R analyses. Similarly, we have also performed some experiments of our own SUBHO modeling using both a Julia and an Rcpp \citep{eddelbuettel2014rcpparmadillo} implementation. We find that these versions are often many times faster than the reported R code implementation times shown  in Table~\ref{Computational Time}. Typically, the Rcpp integration seems to be about seven times faster than the current R code whilst the Julia version, if R integration is used, is a bit slower in comparison to Rcpp. If we do not use any integration and directly use Julia as the compiler, the Horseshoe sampler is roughly five times faster than R. Additionally, all of the Bayesian methods require decisions about MCMC run time (number of iterations, burn-in, etc.) that will  influence the overall computational costs. We have striven to make reasonable choices based on  investigation of standard convergence criteria without regard to  the computational time required, but other users may make different choices when diagnosising  convergence, potentially leading to somewhat different conclusions.

		\section{Further Details on the TNBC Analysis}\label{appendix: tnbc detailed discussion of network}

		\begin{table}
			\centering
			\small
			\caption{Top $10$ Most Connected Genes From Each Method - METABRIC Data}
			\label{table: TNBC top 10 most connected genes}
			\begin{tabular}{*{5}{l}}
				\hline
				\multicolumn{5}{c}{Top Genes Names per Method}\\
				\hline
				\multicolumn{1}{p{1cm}}{Method/ Genes} &
				\multicolumn{1}{p{1cm}}{BASAD AND} &
				\multicolumn{1}{p{1cm}}{SUBHO AND} &
				\multicolumn{1}{p{1cm}}{LASSO AND} &
				\multicolumn{1}{p{1.5cm}}{Hyper-g (BAS)AND} \\
				\hline
				$	1	$	&	FOXC1	&	KIAA1949	&	CSAD	&	FSCN1	\\
				$	2	$	&	SRD5A1	&	FOXN2	&	SSH3	&	ASPM	\\
				$	3	$	&	ATP5G2	&	FAM174A	&	SELENBP1	&	MPZL2	\\
				$	4	$	&	NOP2	&	PIGT	&	LZTFL1	&	LPPR2	\\
				$	5	$	&	MRFAP1L1	&	ANKRD11	&	IRX5	&	DEK	\\
				$	6	$	&	LYAR	&	TSPAN13	&	NCS1	&	HK3	\\
				$	7	$	&	HK3	&	PSME4	&	SUV420H1	&	KRT16	\\
				$	8	$	&	SH3GLB2	&	ITGA3	&	BRD8	&	ROPN1	\\
				$	9	$	&	TMEM115	&	BCL11A	&	CALU	&	FOXA1	\\
				$	10	$	&	HLA-DOB	&	TGFB3	&	PM20D2	&	CCNB2	\\
				\hline
				\multicolumn{1}{p{1cm}}{Method/ Genes} &
				\multicolumn{1}{p{1cm}}{GLASSO} & 
				\multicolumn{1}{p{1.5cm}}{GHS 50\% CI} & 
				\multicolumn{1}{p{1.5cm}}{IW 50\% CI} & 
				\multicolumn{1}{p{1cm}}{Median BDG}  \\
				\hline
				$	1	$	&	TMOD1	&	FAM19A3	&	HORMAD1	&	FOXN2	\\
				$	2	$	&	IGF2BP3	&	SCRG1	&	IL1R2	&	PNP	\\
				$	3	$	&	FABP7	&	GJB3	&	CXorf61	&	ADA	\\
				$	4	$	&	GJB3	&	ARHGEF4	&	MAP2	&	PLCD4	\\
				$	5	$	&	GABBR2	&	PNMA3	&	TMOD1	&	DLGAP5	\\
				$	6	$	&	CXorf61	&	PGBD5	&	GABBR2	&	IGF1R	\\
				$	7	$	&	BX106902	&	RELB	&	CPA4	&	TMEM30B	\\
				$	8	$	&	HORMAD1	&	PM20D2	&	LOXL4	&	CACNA2D2	\\
				$	9	$	&	ATP6V1C2	&	IRX5	&	IGF2BP3	&	ELF4	\\
				$	10	$	&	PTCHD1	&	RNF150	&	CDKN2A	&	ERBB3	\\
				\hline      
				\hline
			\end{tabular}
		\end{table}

		Table~\ref{table: TNBC top 10 most connected genes} reports the names of the top 10 most connected genes from each of the methods, as determined by $G_a$ and $P_a$ (Section~$6.2$). Further analyses and literature review can be conducted based on these identified genes to further  investigate their roles  in breast cancer (BC), generally, and in triple negative breast cancer (TNBC), specifically. While a full investigation is beyond the scope of the current project, we briefly note all of the top $10$ genes for our SUBHO method have previously been reported as playing some role in BC. In the following, we briefly reference a set of key results for each of these top genes.

		\begin{enumerate}
			\item KIAA1949 acts as a suppressor gene in BC \citep{su2010undetectable} .
			\item Downregulation of FOXN2 has been observed in BC tissues \citep{ye2019foxn2}.
			\item FAM174A is among the top-correlated genes of MAGED2, which is known to have  metastatic potential in TNBC  \citep{thakur2022deletion}.
			\item Overexpression of PIGT has been noted in BC cell lines and primary tumors \citep{wu2006overexpression}.
			\item In patients with luminal A ER+  BC, higher expression of ANKRD11 correlates with poor survival \citep{zhou2023serpina3}.
			\item Among BC patients, a lower expression of TSPAN13 was associated with poorer prognosis \citep{wang2019mir}.
			\item PSME4 leads to overactivation of the mTOR signaling pathway \citep{el2018rictor}, a pathway that is often found to be activated in tumors \citep{zou2020mtor}.
			\item Downregulation of ITGA3 has been noted in BC.  Moreover, ITGA3 can promote cell proliferation \citep{zhang2020itga3}.
			\item BCL11A has been found to be upregulated among TNBC patients, and it is a predictor of poor clinical outcomes \citep{wang2020bcl11a}.
			\item High expression of TGFB3 corresponds to significantly worse overall survival in both ER- and basal-like BC \citep{rosas2021positive}.
		\end{enumerate}

\begin{figure} 
			\includegraphics[width=\textwidth]{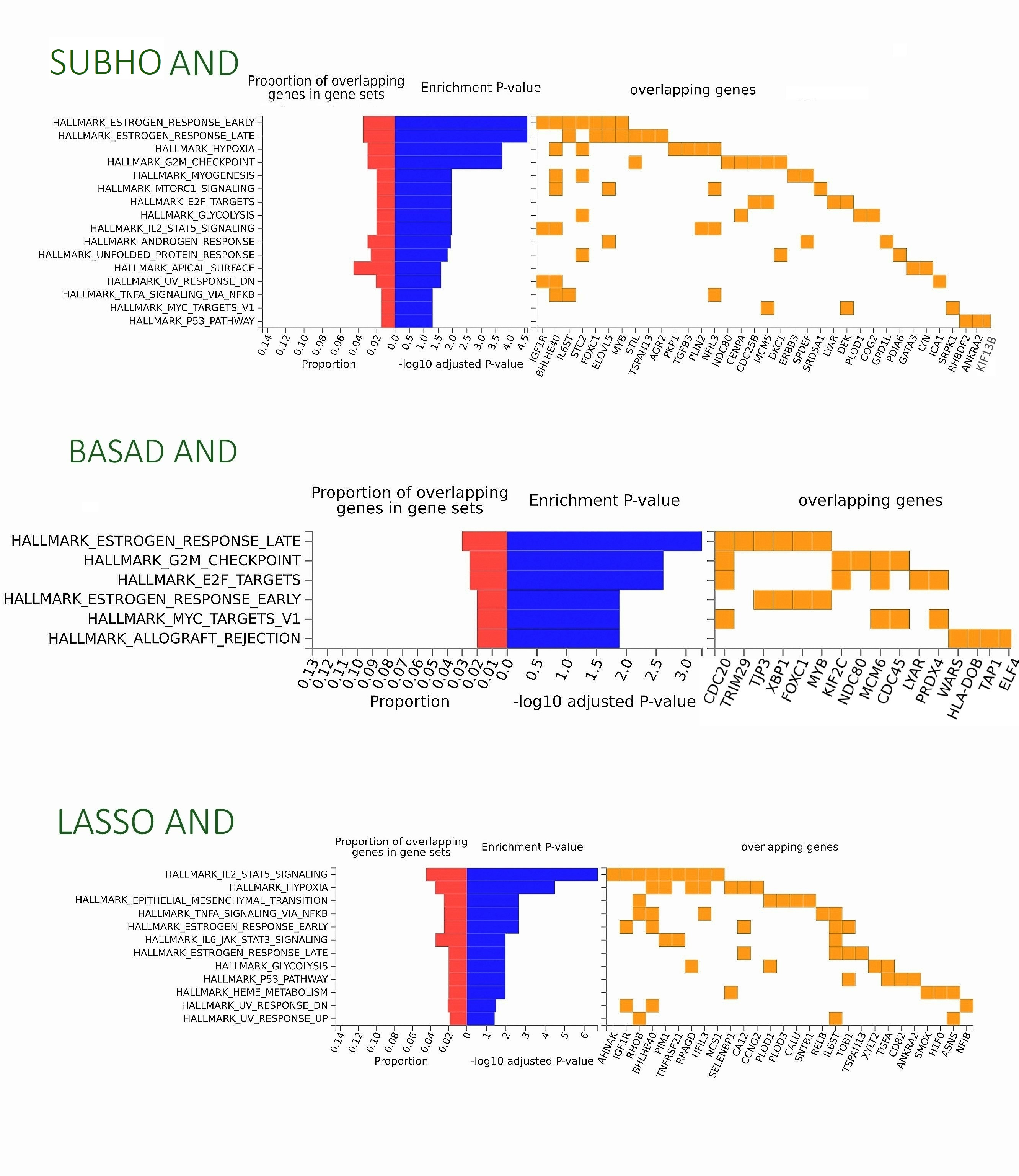}
			\caption{Complete result for significant enrichment with corresponding genes for SUBHO, BASAD and LASSO.}
			\label{fig: TNBCdata_hallmark_withOverlaps 1}
		\end{figure}

\begin{figure} 
			\includegraphics[width=\textwidth]{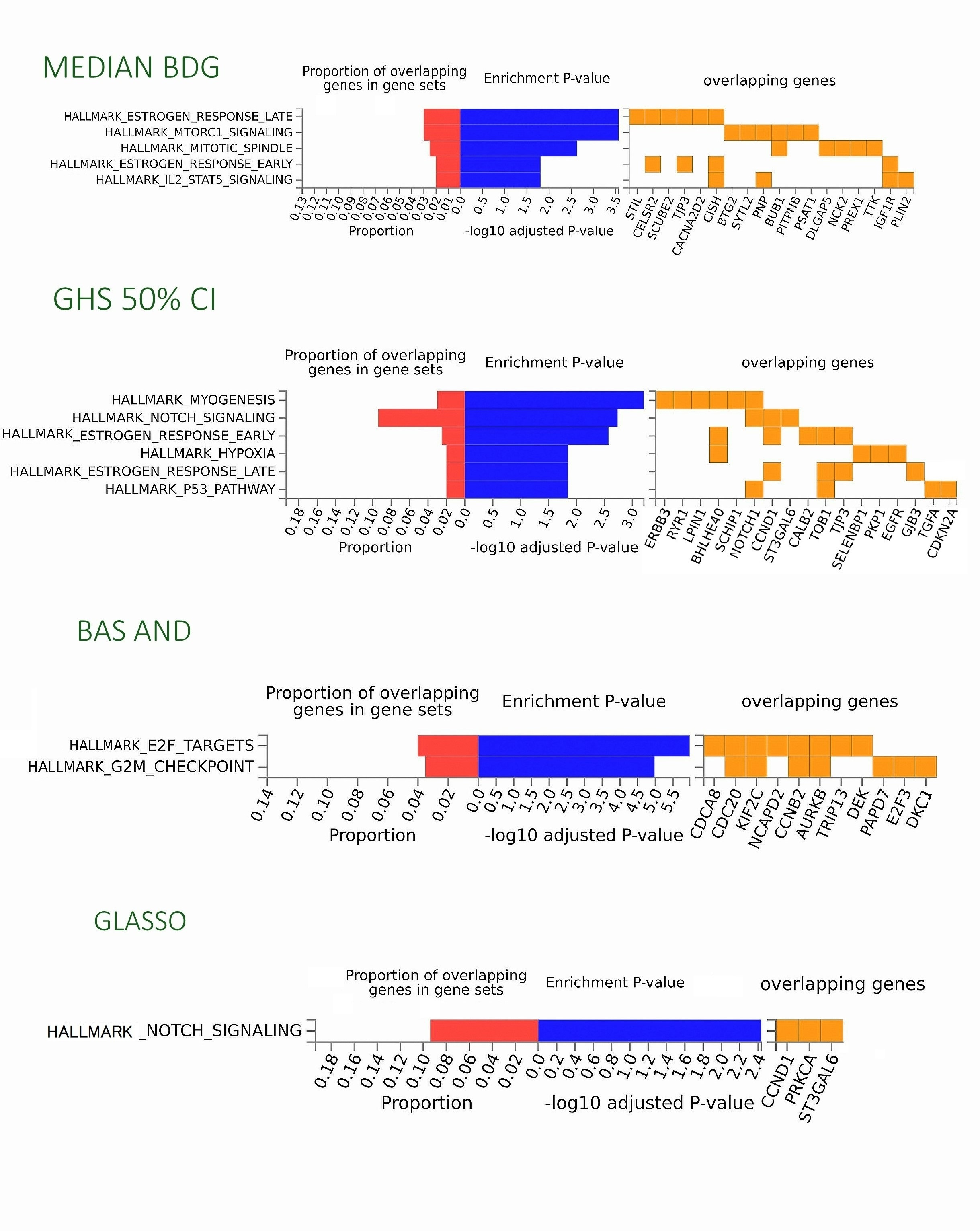}
			\caption{Complete result for significant enrichment with corresponding genes for BDG, GHS, BAS and GLASSO.}
			\label{fig: TNBCdata_hallmark_withOverlaps 2}
		\end{figure}

		Next, we again consider  the set of enrichment results from  FUMA \citep{watanabe2017functional}. As discussed in the main manuscript, these results are based on the  $100$ most connected genes  from the graphs from each method. Figures~\ref{fig: TNBCdata_hallmark_withOverlaps 1} and  \ref{fig: TNBCdata_hallmark_withOverlaps 2} represent an enhanced version of Figure 7 from the manuscript.  In addition to displaying the most enriched hallmark gene sets, this figure also identifies  the individual genes that correspond to each hallmark gene set.

		As with the list of most connected genes from Table~\ref{table: TNBC top 10 most connected genes}, one could consider an in depth investigation based on the set of genes identified from these results. To that end, we note from Figure~\ref{fig: TNBCdata_hallmark_withOverlaps 1} that the SUBHO analysis identifies 34 genes (out of the 100 most connected) to be associated with the 16 significantly enriched hallmark gene sets.
		(See also Table \ref{table: TNBCEnrichment_table}.)
		Investigation of these genes finds that majority of these genes have been cited within the literature as playing some role in the development and/or progression of cancer with the vast majority found to be related to  breast cancer or TNBC.  In Table~\ref{table: References of FUMA Genes}, we provide citations for published results of this sort, listing one each  for 30 of these 34 genes. For the remaining four (COG2, ICA1, ANKRA2, KIF13B), there was no clear evidence that we found of a relationship between these genes and breast cancer. Again, we note that this is not intended to be a comprehensive investigation of the literature around these genes, but our goal is simply to recognize that the most relevant genes from our SUBHO analysis represents a biologically plausible result that is consistent with the published literature.
		
		\begin{table}
			\centering
			\caption{Enrichment from FUMA - METABRIC Data}
			\label{table: TNBCEnrichment_table}
			\begin{tabular}{ccc}
				\hline
				\hline
				\multicolumn{1}{c}{No.} &  
				\multicolumn{1}{c}{Hallmark Gene Sets} &
				\multicolumn{1}{c}{Corresponding Genes}\\
				\hline
				1 & ESTROGEN RESPONSE 	&	IGF1R, BHLHE40, IL6ST, 	\\
				&  EARLY&  STC2, FOXC1, \\
				& & ELOVL5, MYB\\
				2 & ESTROGEN RESPONSE 	&	STIL, IL6ST, FOXC1, \\
				&  LATE &  ELOVL5, MYB, \\
				& & TSPAN13, AGR2\\
				3 &  HYPOXIA	&	PKP1, TGFB3, BHLHE40,  \\
				&  & STC2, PLIN2, NFIL3\\
				4 & G2M CHECKPOINT	&	STIL, NDC80, CENPA, 	\\
				&   & CDC25B, MCM5, DKC1\\
				5 & MYOGENESIS	&	ERBB3, BHLHE40, 	\\
				&  & STC2, SPDEF\\
				6 & MTORC1 SIGNALING	&	BHLHE40, SRD5A1,  \\
				&  &  ELOVL5, NFIL3	\\
				7 & E2F TARGETS	&	CDC25B, MCM5,  \\
				&  &  LYAR, DEK	\\
				8 & GLYCOLYSIS	&	PLOD1, COG2,  \\
				&  &  CENPA, STC2	\\
				9 & IL2 STAT5 	&  IGF1R, BHLHE40, \\
				& SIGNALING   &   PLIN2, NFIL3	\\
				10 & ANDROGEN RESPONSE	&	GPD1L, SPDEF, ELOVL5	\\
				11 & UNFOLDED PROTEIN 	&	PDIA6, STC2, \\	
				&  RESPONSE &  DKC1\\
				12 & APICAL SURFACE	&	GATA3, LYN	\\
				13 & UV RESPONSE DN	&	IGF1R, BHLHE40, ICA1	\\
				14 & TNFA SIGNALING 	&	BHLHE40, IL6ST, \\
				&  VIA NFKB  &  NFIL3	\\
				15 & MYC TARGETS V1	&	MCM5, DEK, SRPK1	\\
				16 & P53 PATHWAY	&	RHBDF2, ANKRA2, KIF13B	\\
				\hline
			\end{tabular}
		\end{table}

		\begin{table}
			\centering
			\caption{References for SUBHO-TNBC Genes in Table~\ref{table: TNBCEnrichment_table}}
			\label{table: References of FUMA Genes}
			\begin{tabular}{ccc}
				\hline
				\hline
				\multicolumn{1}{c}{Fig.~\ref{fig: TNBCdata_hallmark_withOverlaps 1}}   \\
				\multicolumn{1}{c}{Position} &
				\multicolumn{1}{c}{Gene Name} &
				\multicolumn{1}{c}{Reference} \\
				\hline
				1	&	IGF1R	&	\cite{li2019structural,obr2022breast}			\\
				2	&	BHLHE40	&	\cite{devaux2020long}			\\
				3	&	IL6ST	&	\cite{martinez2021signal}			\\
				4	&	STC2	&	\cite{qie2022stanniocalcin}       			\\
				5	&	FOXC1	&	    \cite{elian2018foxc1}   			\\
				6	&	ELOVL5	&	   \cite{kieu2022downregulation}			\\
				7	&	MYB	& 	     \cite{ciciromyb}   			\\
				8	&	STIL	&	    \cite{kolberg2022impact}      			\\
				9	&	TSPAN13	&	   \cite{jiang2019expression}			\\
				10	&	AGR2	&	    \cite{salmans2013estrogen}\\
				11	&	PKP1	&	    \cite{li2021desmosomal}    			\\
				12	&	TGFB3	&	      \cite{vishnubalaji2021epigenetic}			\\
				13	&	PLIN2	&	     \cite{he2022lipid}  			\\
				14	&	NFIL3	&	     \cite{yang2022elevated}     			\\
				15	&	NDC80	&	   \cite{li2022ndc80}        			\\
				16	&	CENPA	&	   \cite{qiu2013prognostic}      			\\
				17	&	CDC25B	&	   \cite{cairns2020cdc25b}     			\\
				18	&	MCM5	&	   \cite{sheng2021lncrna}   			\\
				19	&	DKC1	&	   \cite{guerrieri2020dkc1}      			\\
				20	&	ERBB3	&	  \cite{uliano2023targeting}    			\\
				21	&	SPDEF	&	   \cite{ye2020double}  			\\
				22	&	SRD5A1	&	   \cite{zhang2021over}   			\\
				23	&	LYAR	&	    \cite{chen2021analysis}    			\\
				24	&	DEK	&	   \cite{li2017association}  			\\
				25	&	PLOD1	&	   \cite{li2017association}     			\\
				27	&	GPD1L	&	  \cite{zhou2017identification}			\\
				28	&	PDIA6	&	   \cite{yang2022roles}     			\\
				29	&	GATA3	&	   \cite{hruschka2020gata3}     			\\
				31	&	SRPK1	&	    \cite{malvi2020limk2}      			\\
				32	&	RHBDF2	&	    \cite{masood2022investigating}			\\
				\hline
				26	&	COG2	&	Relationship to BC unknown\\
				30	&	ICA1	&   Relationship to BC unknown\\
				33	&	ANKRA2	&   Relationship to BC unknown\\
				34	&	KIF13B	&	Relationship to BC unknown\\ 
				\hline \hline
			\end{tabular}
		\end{table}
		
		\clearpage

		\section{Dutch Breast Cancer Data Analysis}\label{appendix: Dutch BC data}
		
		We further illustrate all the methods with a real data example taken from \cite{atchade2019quasi} and discussed in \cite{van2002gene}. Gene expression profiles were obtained at diagnosis for 295 women with breast cancer at the Nederlands Kanker Instituut in Amsterdam, Netherlands. Patients were followed over time and are grouped  based on whether the cancer meatstasized within five years.	Of the $295$ patients, $101$ developed a metastasis within the first five years.	As mentioned in \cite{atchade2019quasi}, the initial data set had $24,884$ genes. To reduce the data dimension, marginal logistic regression of metastatis (occured or not) on each gene were performed, and only those genes were selected for which the $p$-value was smaller than $0.005$. A total of $812$ genes were thus selected since these were the genes that are deferentially expressed by metastasis status. We limit further analysis to the $n=101$ patients with metastasis, and we seek to learn the genetic expression network among the $p=812$ identified genes. The expression data was previously normalized, thus we center and re-scale the gene expressions after subsetting before our analyses. We implement all methods previously considered in the simulation study.  We note here that the slightly smaller $p$ and the substantially smaller $n$ has a major impact on the computational feasibility of the LOOCV step within the Projpred method. Unlike the TBNC analysis, we are able to include their method here.

		\begin{figure}
			\includegraphics[width=\textwidth]{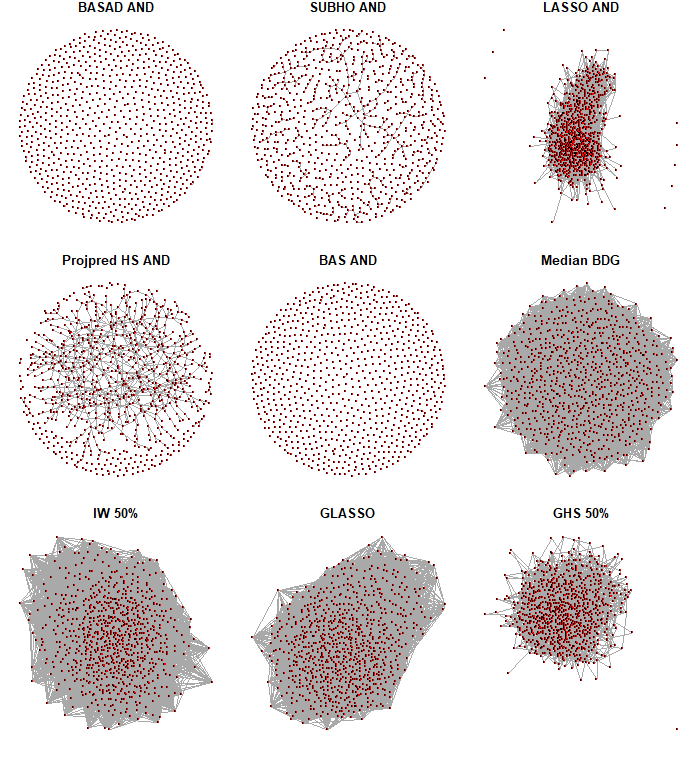}   
			\caption{Gene networks from Dutch Breast Cancer data } 
			\label{fig:data_gene_network}
		\end{figure}

		We discuss the estimated Gene Networks shown in Figure~\ref{fig:data_gene_network}.  There are $329,266$ possible edges that could be included in the graphs, but as before, we anticipate that the underlying network should be fairly sparse. First, we can see the estimated graph from the BASAD AND ($34$ estimated edges) at the top left contains highly disjoint sets of nodes with very few connections between nodes. Moving to the right, the estimated graph from the SUBHO AND ($512$ edges) which contains a number of chain-like structures connecting small collections of associated genes. Next, we have LASSO AND ($3726$ edges), which shows a few singleton nodes towards the edges, but the graph is dominated by a large collection  densely connected nodes. In the second row, Projpred HS AND ($812$ edges) shows a network that is a bit denser than SUBHO network with a much more connected network in the center of graphic. BAS (hyper-$g$) produces an extreme sparse graph ($37$ edges), similar to BASAD, whereas BDG ($14,101$ edges), IW ($60,883$ edges), GLASSO ($19,891$ edges) and GHS ($2814$ edges) all produce graphs that are substantially more dense.

\begin{figure}
			\centering
			\includegraphics[width=\textwidth]{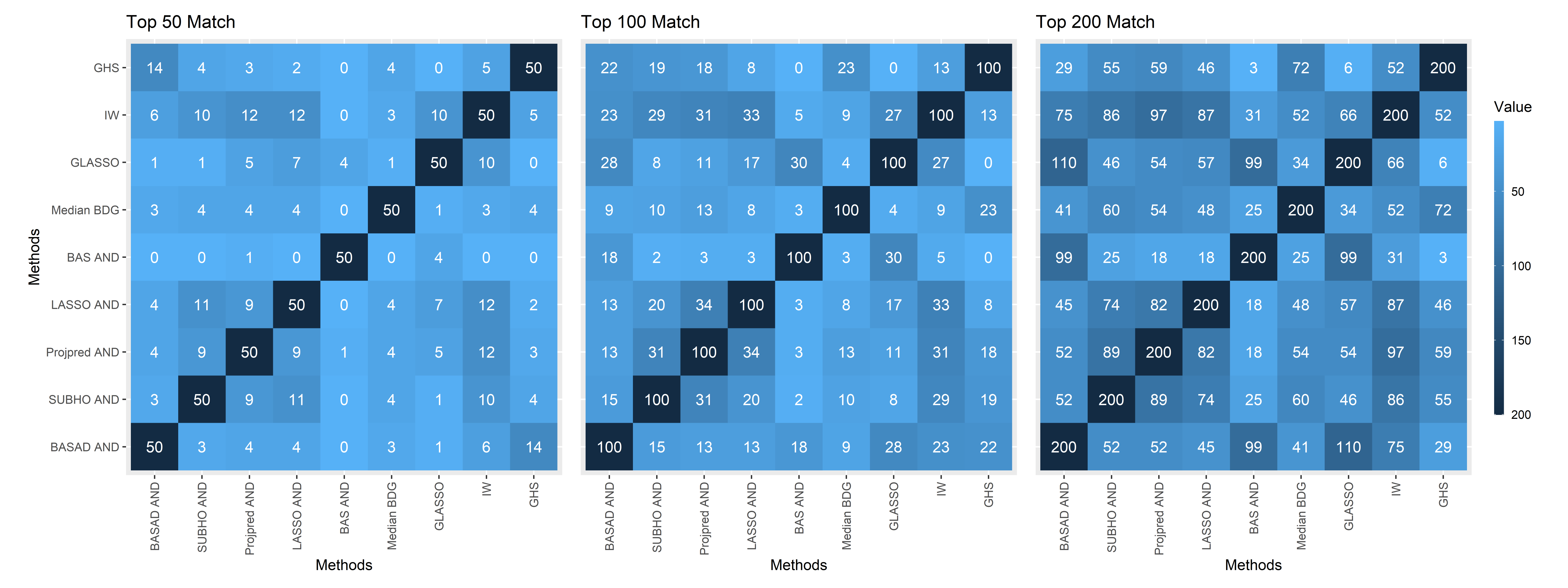} 
			\caption{Comparison of Gene Name Match}
			\label{fig: Atchade Gene name mis-match}
		\end{figure}

		We follow the same strategy as in Section 6.2 to identify the most connected genes in each of the estimated networks. The similarities across the various methods in terms of the top $K\in\{50,100,200\}$ most connected genes are  displayed in  Figure~\ref{fig: Atchade Gene name mis-match}. The main  takeaway is that majority of methods are finding very distinct sets of top genes with relatively little overlap across the different estimation approaches. 
		
		In Table~\ref{table: top 10 most connected genes} we report the top ten most connected genes from each method. We note from this list that many of these identified genes are actually affymetrix probe IDs. Such probes are used to measure the expression level of many genes. In general, before conducting any downstream analysis of a microarray experiments, probe IDs are converted into Genes (ensembl ID or entrez ID). There exist some tools for conversion of the underlying mapping of probes to bioentities (i.e., genes or proteins) such as DAVID \citep{barrett2012ncbi}. We embarked upon this route, but more than $50\%$ of these probe IDs did not  match with any known genes. 
		As more than half of these genes in our analysis cannot be matched to known gene names, we do not attempt to perform further investigation such as the FUMA analysis that we considered in the TNBC data.

		\begin{table}
			\tiny 
			\centering
			\caption{Top $10$ Most Connected Genes for Dutch Breast Cancer Data}
			\label{table: top 10 most connected genes}
			\begin{tabular}{*{10}{l}}
				\hline
				\hline
				\multicolumn{1}{p{1cm}}{Method/ Genes} &
				\multicolumn{1}{p{1cm}}{BASAD AND} &
				\multicolumn{1}{p{1cm}}{SUBHO AND} &
				\multicolumn{1}{p{1cm}}{LASSO AND} &
				\multicolumn{1}{p{1cm}}{BAS AND} &
				\multicolumn{1}{p{1cm}}{Projpred HS AND} \\
				\hline
				$	1	$	&	$	Contig31646_RC	$	&	$	Contig57023_RC	$	&	$	NM_002706	$	&	$	Contig45574	$	&	$	AF100756	$	\\
				$	2	$	&	$	Contig48867_RC	$	&	$	AF007217	$	&	$	Contig48004_RC	$	&	$	AB037770	$	&	$	Contig3902_RC	$	\\
				$	3	$	&	$	Contig33834_RC	$	&	$	AF100756	$	&	$	NM_001130	$	&	$	Contig52786_RC	$	&	$	Contig50388_RC	$	\\
				$	4	$	&	$	AL049415	$	&	$	AL050148	$	&	$	AF279865	$	&	$	Contig49175_RC	$	&	$	AB007883	$	\\
				$	5	$	&	$	Contig44289_RC	$	&	$	NM_007204	$	&	$	NM_014174	$	&	$	NM_006607	$	&	$	Contig55377_RC	$	\\
				$	6	$	&	$	NM_018460	$	&	$	AF146760	$	&	$	AL122091	$	&	$	NM_014359	$	&	$	Contig51442_RC	$	\\
				$	7	$	&	$	AF035284	$	&	$	AB007883	$	&	$	Contig55421_RC	$	&	$	NM_019110	$	&	$	Contig66868_RC	$	\\
				$	8	$	&	$	NM_006096	$	&	$	NM_017704	$	&	$	NM_006579	$	&	$	NM_000717	$	&	$	AF161415	$	\\
				$	9	$	&	$	NM_018454	$	&	$	Contig41413_RC	$	&	$	AL050064	$	&	$	U75968	$	&	$	NM_017704	$	\\
				$	10	$	&	$	AB007969	$	&	$	Contig43621_RC	$	&	$	NM_005320	$	&	$	NM_004399	$	&	$	AB007969	$	\\
				\hline
				\multicolumn{1}{p{1cm}}{Method/ Genes} &
				\multicolumn{1}{p{1cm}}{GLASSO} & 
				\multicolumn{1}{p{1cm}}{GHS 50\% C.I} & 
				\multicolumn{1}{p{1cm}}{InvWish 50\% C.I} & 
				\multicolumn{1}{p{1cm}}{Median BDG}  & &  \\
				\hline
				$	1	$	&	$	AL117442	$	&	$	NM_002036	$	&	$	NM_005252	$	&	$	Contig19153_RC	$	\\
				$	2	$	&	$	NM_001109	$	&	$	NM_006579	$	&	$	AF073299	$	&	$	Contig42026_RC	$	\\
				$	3	$	&	$	NM_018373	$	&	$	NM_000125	$	&	$	AL137449	$	&	$	Contig48004_RC	$	\\
				$	4	$	&	$	NM_007063	$	&	$	Contig3902_RC	$	&	$	NM_005165	$	&	$	NM_018287	$	\\
				$	5	$	&	$	Contig36106_RC	$	&	$	AL080095	$	&	$	Contig36448_RC	$	&	$	AL050064	$	\\
				$	6	$	&	$	NM_007278	$	&	$	NM_001130	$	&	$	Contig48328_RC	$	&	$	NM_003176	$	\\
				$	7	$	&	$	NM_014174	$	&	$	AF007217	$	&	$	NM_006765	$	&	$	Contig45574	$	\\
				$	8	$	&	$	NM_012247	$	&	$	Contig50814_RC	$	&	$	Contig42011_RC	$	&	$	NM_000365	$	\\
				$	9	$	&	$	AL122091	$	&	$	NM_001540	$	&	$	Contig38580_RC	$	&	$	NM_004780	$	\\
				$	10	$	&	$	AF015041	$	&	$	Contig57023_RC	$	&	$	Contig27749_RC	$	&	$	Contig54968_RC	$	\\
				\hline
			\end{tabular}
		\end{table}

		\bibliographystyle{ba}
		\bibliography{Fast_Bayesian_High_Dimensional_Gaussian_Graphical_Model_Estimation}
		
		\begin{acks}[Acknowledgments]
			This research was partially supported by the U.S. NSF under grant CNS1828521. We used computational resources of the University of Louisville Cardinal Research Cluster. We acknowledge the help with data from Dr.~Diptavo Dutta, Investigator DCEG, NCI.
		\end{acks}
		
	\end{document}